\documentclass[aps,prb,twocolumn,groupedaddress]{revtex4-1}

\usepackage[utf8]{inputenc}
\usepackage[]{amsmath}
\usepackage{amsfonts}
\usepackage{amssymb}
\usepackage[]{color}
\usepackage[english]{babel}

\usepackage{graphicx}

\usepackage {nicefrac}
\usepackage{todonotes} 

\usepackage{tikz}
\usepackage{pgfplots}
\pgfplotsset{width=7cm}

\renewcommand{\vec}[1]{\boldsymbol #1}

\begin{document}


\title{Functional renormalization group for fermion lattice models in three dimensions: application to the Hubbard model on the cubic lattice}


\author{J. Ehrlich}
\affiliation{Institute for Theoretical Solid State Physics, RWTH Aachen University, Aachen, Germany}
\affiliation{Peter Gr\"unberg Institute and Institute for Advances Simulation, Forschungszentrum J\"ulich, Germany}
\email[]{j.ehrlich@fz-juelich.de}
\author{C. Honerkamp}
\affiliation{Institute for Theoretical Solid State Physics, RWTH Aachen University, Aachen, Germany}
\affiliation{JARA-FIT, J\"ulich-Aachen Research Alliance - Fundamentals of Future Information Technology, Germany}


\date{\today}

\begin{abstract}
The channel-decomposed functional renormalization group (FRG) approach, most recently in the variant of truncated-unity-(TU-)FRG, has so far been used for various two-dimensional model systems. Yet, for many interesting material systems the third spatial dimension is of clear relevance. Therefore FRG schemes working in three spatial dimensions (3D) are definitely on the wishlist. Here we demonstrate that a 3D TUFRG scheme can be set up in straightforward extension of previous 2D codes and gives physically sensible results with affordable numerical effort, both regarding the qualitative as well as the quantitative description. The computed phase diagram of the three-dimensional Hubbard model at half filling or perfect nesting shows a phase transition to a \((\pi,\pi,\pi)\)-ordered antiferromagnetic ground state for repulsive interactions at an energy scale that compares well with other numerical approaches in the literature. Furthermore, the method allowed us to detect a \(d\)-wave pairing and a concurring \((\pi,\pi,0)\) antiferromagnetic ground state in the hole doped Hubbard model.
\end{abstract}


\maketitle


\section{Introduction}
Quantum materials with significant correlation effects arising from the interactions within the electron system are under intense experimental and theoretical studies, as they give rise to a plethora of interesting phases like magnetic orderings and unconventional superconductivity, with potential uses in future technologies.
One of the simplest lattice models leading to correlation effects is the single-band Hubbard model, where electrons interact via a repulsive coupling \(U\) with each other only when they are located at the same lattice site. Despite its simplicity, analytic solutions of this model for \(U\not= 0\) exist only in the limit of one\cite{} and infinite dimensions\cite{Georges1992}.\\
The Hubbard model exhibits in the case of three spatial dimensions and half-filling a transition to an anti-ferromagnetic (AFM) ordered ground state for all interaction strength \(U\) at finite N\'eel temperatures \(T_\text{N}\). This \(T_\text{N}\) depends on the interaction strength \(U\), as the transition is attributed to different effects: In the weak interaction limit it is caused by thermal spin-flip excitations across the Fermi surface, which can be described resonably well by the random phase approximation (RPA) extended by corrections due to local quantum fluctuations \cite{Hirsch1987,Rodero1992,Freericks1995}. In the strong coupling case the phase transition is caused by spin-spin interactions of local moments. Thus, the Hubbard model converts to the Heisenberg model with \(J=t^2/U\) and the N\'eel temperature tends to the Heisenberg limit \(3.83/U\)\cite{Sandvik1998,Affleck1988}.
In this limit, when the interaction reaches the order of the bandwidth, a Mott-Insulator transition (MIT) can be detected above the magnetic ordering temperature \cite{Staudt2000},  maintaining an excitation gap for the magnetically disordered phase at larger \(U\). \\
As the \(T_\text{N}(U)\)-curve grows from these limiting cases towards intermediate coupling strength, it is expected to exhibit a maximum in this region, which was shown to go along a pseudogap-like behavior\cite{Fuchs2011}. But as no analytic solution exists, the exact form of \(T_\text{N}(U)\) has been targeted by a numerous methods\cite{Hirschmeier2015,Staudt2000,Kent2005,Kozik2013,Rohringer2011}. Therefore, the N\'eel curve \(T_\text{N}(U)\) of the three dimensional half-filled Hubbard model can be used as a reference for new methods.\\ 
In addition to this specific case, the three dimensional Hubbard model has been extended and studied in different specific aspects. With an additional next-nearest neighbor hopping \(t'\), the magnetic order is reduced due to magnetic frustration\cite{Fuchs2011}. If, on the other hand, nearest neighbor interactions are included the Hubbard model exhibits a rich phase diagram of different magnetic orders, charge and spin density waves and superconductivity\cite{Dongen1991,Dongen1994}. The simplest way, doping the system away from half-filling, showed that \(d\)-wave pairing can become dominant\cite{Scalapino1986}. This richness of phases occurring for sometimes already small detunings away from perfect nesting makes it important to treat all contributions to correlation effects on the same footing.\\
The functional renormalization group (FRG) appears to be a suitable method in this regard, as it treats all three channels (spin, charge and pairing) on the same footing. Although the FRG is well established for investigations in one- or two-dimensional correlated systems\cite{Kopietz2010,Metzner2012,Platt2013}, it has not been applied to lattice systems in three dimensions so far due to numerical costs of the most often used \(N\)-patch Brillouin zone discretization scheme. In recent years however, channel-decomposition techniques in conjunction  with form-factor expansions have defined a new and powerful FRG line of approach\cite{Husemann2009,Husemann2012,Eberlein2015,Wang2012,Wang2013,Lichtenstein2017,Schober2018}. In contrast to the \(N\)-patch schemes, a much higher momentum resolution can be reached in the new schemes\cite{Lichtenstein2017,Sanchez2017} and various other improvements like a detailed frequency dependence of the interactions\cite{Vilardi2017} could be built in. Most notably, in a recent study, multi-loop corrections were incorporated and shown to converge for the 2D Hubbard model at perfect nesting\cite{Tagliavini2019}. Thus the method can be understood as quantitatively reliable, at least for simple situations. A neat way to derive the FRG equations for in this new FRG approach is the insertion of truncated resolutions of unity into the loop diagrams. This allows to cast the flow equations for the interaction vertex in the form of matrix products which can be parallelized efficiently. This technique is known as truncated-unity-FRG (TUFRG\cite{Lichtenstein2017} ) and will be employed in what follows. On the level used here, the TUFRG can be viewed as a reformulation of the SMFRG by Wang et al.\cite{Wang2012,Wang2013}. \\
The numerical cost of the TUFRG scales quadratically in the number of Brillouin zone points taken into account and quartically with the number of form-factors kept. If we now extend the scheme from 2 to 3 dimensions while maintaining a regular grid, the number of BZ points increases from \(n_\text{kpt}^2\) to \(n_\text{kpt}^3\), but this may be still tolerable if we do not need ultra-high precision. The number of form factors, however, increases but remains comparable, thus making a treatment of three-dimensional models by TUFRG possible. In this paper we, therefore, apply it to the three dimensional Hubbard model, while neglecting the self-energy flow (which can be built in in subsequent works along the lines of Ref. \onlinecite{Tagliavini2019}). By this we reproduce the weak-coupling part of the AFM phase diagram, and find  \(d\)-wave pairing in the doped three-dimensional Hubbard model.\\
This paper is organized as follows. In section \ref{sec:Method} we shortly introduce the truncated-unity functional renormalization group equations and describe its implementation as well as the model in section \ref{sec:Model}. In section \ref{res:phasediag} we discuss the phase diagram and in section \ref{res:doped} we show the arising \(d\)-wave pairing when doping the Hubbard model, before we conclude in section \ref{conclusion}. 

\section{Method} \label{sec:Method}
\begin{figure}
  \includegraphics[width=0.9\columnwidth]{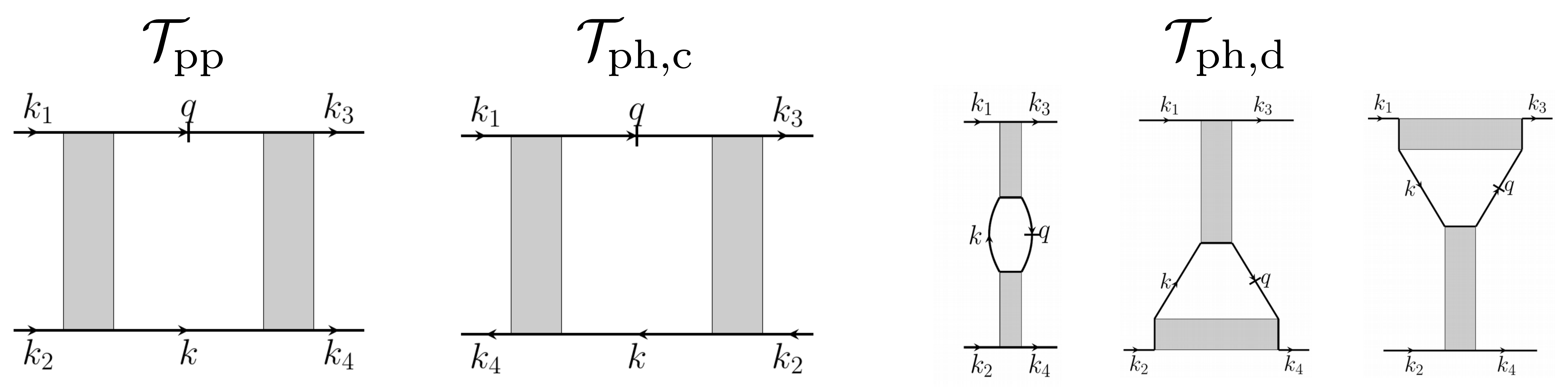}
  \caption{Diagrammatic representation of the 1-PI functional RG interaction at one-loop truncation, representing the particle-particle (\(\mathcal{T}_\text{pp}\)), the crossing  (\(\mathcal{T}_{\text{ph, c}}\)) and the direct (\(\mathcal{T}_{\text{ph, d}}\)) partticle-hole contributions.}
  \label{frg_diag}
\end{figure}
The truncated-unity functional renormalization group (TUFRG), which we use here, is a recently developed \cite{Lichtenstein2017} efficient approximation of the FRG and, by now, a well-established tool for the investigation of two-dimensional correlated systems\cite{Metzner2012,Tagliavini2019}. 
The basic idea of the one-particle irreducible (1-PI) FRG is to introduce a dependence on a scale parameter \(\Lambda\) to the non-interacting Green's function. Variation of \(\Lambda\) induces a flow of correlation and vertex functions of the system. A typical choice is to implement the scale parameter \(\Lambda\) as an infrared energy or frequency cutoff, such that only modes with absolute value of the band energy or Matsubara frequency larger than the cutoff are considered. Below we actually apply a smoothened frequency cutoff function as first employed by Husemann and Salmhofer\cite{Husemann2009}. In any case, in the limit \(\Lambda\rightarrow 0\) the full system is obtained again. An infinite set of differential flow equations for the effective, \(\Lambda\)-dependent \(n\)-particle interactions can be derived by the \(\Lambda\) derivative of the corresponding generating functional\cite{Salmhofer2001, Metzner2012}. For the case of weak to intermediate interactions, it is usually assumed that it is sufficient to neglect interactions of order larger than \(n\geq3\), drastically simplifying the calculations\cite{Salmhofer2001}. Note that recently, so-called multiloop extensions have been implemented in the FRG\cite{Kugler2018,Tagliavini2019}. These allow one to reduce the error from this truncation to such a degree that equivalence to the parquet approximation is obtained.  \\
As the Hubbard model is SU(2) symmetric and a spontaneous symmetry breaking is only allowed to be approached but not to occur in the present manuscript, we employ the SU(2)-symmetric form of the FRG equations as e.g.\ in ref.\ \onlinecite{Metzner2012}. In addition, we simplify our calculation by neglecting self-energy corrections, such that our main element of interest is the 1-PI interacting vertex \(V\) depending on four momenta, frequencies and orbitals or bands. Exploiting energy and momentum conservation within the vertex reduces the number of frequency and momentum arguments by one. In the flow equation for \(V\) all possible topologically non-equivalent combinations of two vertices connected by a full green function \(G^\Lambda\) and a single scale propagator \(S^\Lambda=-G^\Lambda \left(\frac{\mathrm{d}}{\mathrm{d}\Lambda}G_0^\Lambda\right) G^\Lambda\) are contained. The corresponding contributions, shown in Fig.\,\ref{frg_diag}, can be identified as particle-particle (pp), crossed particle-hole (ph,cr) and direct particle-hole (ph,d) channel.
The numerical effort scales with the fourth order in frequencies and momenta due to the external legs and the internal loop summation, such that an investigation with high resolution is limited by both, memory and compute time. This limitation can be overcome for the momentum part by the TUFRG, which builds on earlier ideas of channel decomposition and form-factor expansion\cite{Husemann2009,Husemann2012} and represents a reformulation of the singular-mode (SM)FRG by Wang et al.\cite{Wang2012,Wang2013}.\\
In the following we sketch the main elements of the TUFRG. For more details, see Refs. \onlinecite{Lichtenstein2017,Schober2018,Tagliavini2019}. In a first step, the full flowing interaction vertex can be split up into four parts, corresponding to the initial interaction and the three channels.
We write
\begin{align}
\begin{split}
V^\Lambda (\vec{k}_1,&\vec{k}_2;\vec{k}_3)=\\
&V_0(\vec{k}_1,\vec{k}_2;\vec{k}_3) - \Phi^{\Lambda, \text{P}}(\vec{k}_1+\vec{k}_2;\vec{k}_1,\vec{k}_3)+\\
&\Phi^{\Lambda, \text{C}}(\vec{k}_3-\vec{k}_2;\vec{k}_1,\vec{k}_3)+\Phi^{\Lambda, \text{D}}(\vec{k}_1-\vec{k}_3;\vec{k}_1,\vec{k}_4) \, .
\end{split}
\end{align} 
Each channel $ \Phi^{\Lambda, \text{X}}$ with $X= P,C,D$  has a characteristic main total momentum or transfer
\begin{align}
\begin{split}
 \vec{s} & \equiv \vec{k}_1+\vec{k}_2 \qquad\text{(pp)}\,, \\[5pt]
 \vec{t} & \equiv \vec{k}_1-\vec{k}_3 \qquad\text{(ph,cr)} \,, \\[5pt]
 \vec{u} & \equiv \vec{k}_2-\vec{k}_3 \qquad\text{(ph,d)}
\end{split}
\end{align}
as first argument. From general arguments and previous calculations (see, e.g.\ ref.\ \onlinecite{Metzner2012} and references therein), a strong dependence on \(\vec{s}\), \(\vec{t}\) or \(\vec{u}\) is known to show up in the full vertex if spontaneous symmetry breaking corresponding to this channel is approached at low scales. 
For the FRG flow, each \(\Phi^{\Lambda, \text{X}} \)'s \(\Lambda\)-derivative is taken to be the corresponding diagrams \(\mathcal{T}^\text{X}\) in fig.\ \ref{frg_diag} which also depend strongly on one of the characteristic momenta \(\vec{s}\), \(\vec{t}\) or \(\vec{u}\).
Besides the strong dependence on \(\vec{s}\), \(\vec{t}\) or \(\vec{u}\), the channel couplings \( \Phi^{\Lambda, \text{X}}\) are found to only weakly depend on their second and third argument. Thus, the weak dependencies can be described\cite{Lichtenstein2017} by a small number of slowly varying basis function in the Brillouin zone, called form-factors. Therefore, in a second step, the weak dependencies are projected to a form-factor basis \(\{f_i\}\) and we call the projections of the \(\Phi^{\Lambda, \-text{X}}\) to this basis the propagator of the X-channel, denoted by \(P^\Lambda_{m,n}(\vec{s})\), \(C^\Lambda_{m,n}(\vec {u})\) and \(D^\Lambda_{m,n}(\vec{t})\). By this procedure each channel propagator has its own flow equation,
\begin{align}\label{TUFRG_eq}
\begin{split}
 \dot{P}^\Lambda_{m,n}(\vec{s}) & = V^{\Lambda,\text{P}}_{m,i}(\vec{s})\chi^{\Lambda,\text{pp}}_{i,j}(\vec{s}) V^{\Lambda,\text{P}}_{j,n}(\vec{s}) \,, \\[5pt]
 \dot{C}^\Lambda_{m,n}(\vec u) & = V^{\Lambda,\text{C}}_{m,i}(\vec  u)\chi^{\Lambda,\text{ph}}_{i,j}(\vec  u) V^{\Lambda,\text{C}}_{j,n}(\vec  u) \,, \\[5pt]
 \dot{D}^\Lambda_{m,n}(\vec t) & = 2 V^{\Lambda,\text{D}}_{m,i}(\vec  t)\chi^{\Lambda,\text{ph}}_{i,j}(\vec  t) V^{\Lambda,\text{D}}_{j,n}(\vec  t) \\ 
                          & + V^{\Lambda,\text{C}}_{m,i}(\vec t)\chi^{\Lambda,\text{ph}}_{i,j}(\vec  t) V^{\Lambda,\text{D}}_{j,n}(\vec t) \\
                          & + V^{\Lambda,\text{D}}_{m,i}(\vec t)\chi^{\Lambda,\text{ph}}_{i,j}(\vec t) V^{\Lambda,\text{C}}_{j,n}(\vec t) \\ 
\end{split}
\end{align}
where \(i\) and \(j\) are form-factor indices to be summed over. These equations are products of three matrices in the form-factor basis, scaling linearly with the number of momenta. The terms
\begin{align}\label{bubble_proj}
\begin{split}
 \chi^{\Lambda,\text{pp}}_{i,j}(\vec{s})=\int \frac{d\vec{q}}{(2\pi)^3} \, f_i^*(\vec{q})f_j(\vec{q})L_\text{pp}(\vec{q},\vec{s})\\
 \chi^{\Lambda,\text{ph}}_{i,j}(\vec{t}|\vec{u})=\int \frac{d\vec{q}}{(2\pi)^3} \, 
 f_i^*(\vec{q})f_j(\vec{q})L_\text{ph}(\vec{q},\vec{t}|\vec{u})
\end{split}
\end{align}
are the projections of the particle-particle and particle-hole bubbles 
 to the form-factor basis. The terms \(V^{\Lambda,\text{X}}_{i,j}\) are the projections of the full vertex to channel X, e.g.
\begin{align}\label{coupleP_proj}
V^{\Lambda,\text{P}}_{i,j} (s) = \int \frac{d\vec{q}}{(2\pi)^3} \int \frac{d\vec{q}'}{(2\pi)^3} \, f_i^*(\vec{q}) V^\Lambda ( \vec{q},-\vec{q}+\vec{s} , \vec{q}') f_j(\vec{q}') \, . 
\end{align} 
 Due to the decomposition of the full vertex \(V^\Lambda\) this requires the projection of the other propagators and the initial interaction to the corresponding channel. For example, projecting the C-channel to the P-channel becomes 
\begin{align}\label{ctop_proj}
\begin{split}
 \hat{P}[\Phi^{\text{C}}]_{m,n}(\vec{s}) = & \int \frac{\mathrm{d}\vec{q}}{(-2\pi)^3}  \frac{\mathrm{d}\vec{q}'}{(2\pi)^3}  \, f^*_m(\vec{q}) f_n(\vec{q}')\\[5pt]
 & \sum_{i,j}f_i(\vec{q})f^*_j(\vec{q}')C_{i,j}(\vec{q}+\vec{q}'-\vec{s})
\end{split}
\end{align}
where the back-projection from form-factor to three-momentum space was used for the C-channel propagator. To conclude, in the TUFRG we have to solve the propagator flow equations \eqref{TUFRG_eq} which require a form-factor projection of the two-propagator term \eqref{bubble_proj} and the channel-to-channel projections like thes one in \eqref{ctop_proj}, which scales linearly with the number of momenta. For a further discussion of the formalism and additional aspects like self-energies, frequency dependence, multi-loop extensions and susceptibilities, see e.g.\ ref.\ \onlinecite{Tagliavini2019}.

\section{Model and implementation details} \label{sec:Model}
\begin{figure}
	\includegraphics[width=0.65\columnwidth]{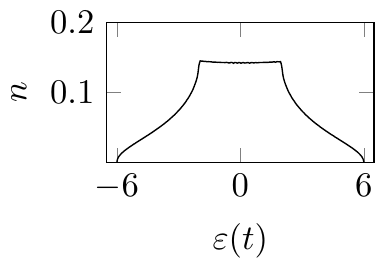}
	\includegraphics[width=0.45\columnwidth]{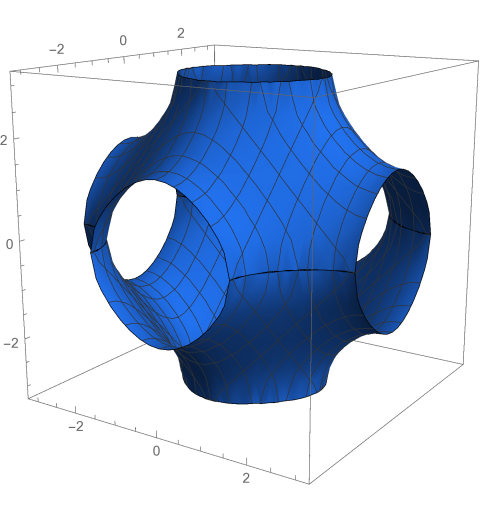}
	\includegraphics[width=0.45\columnwidth]{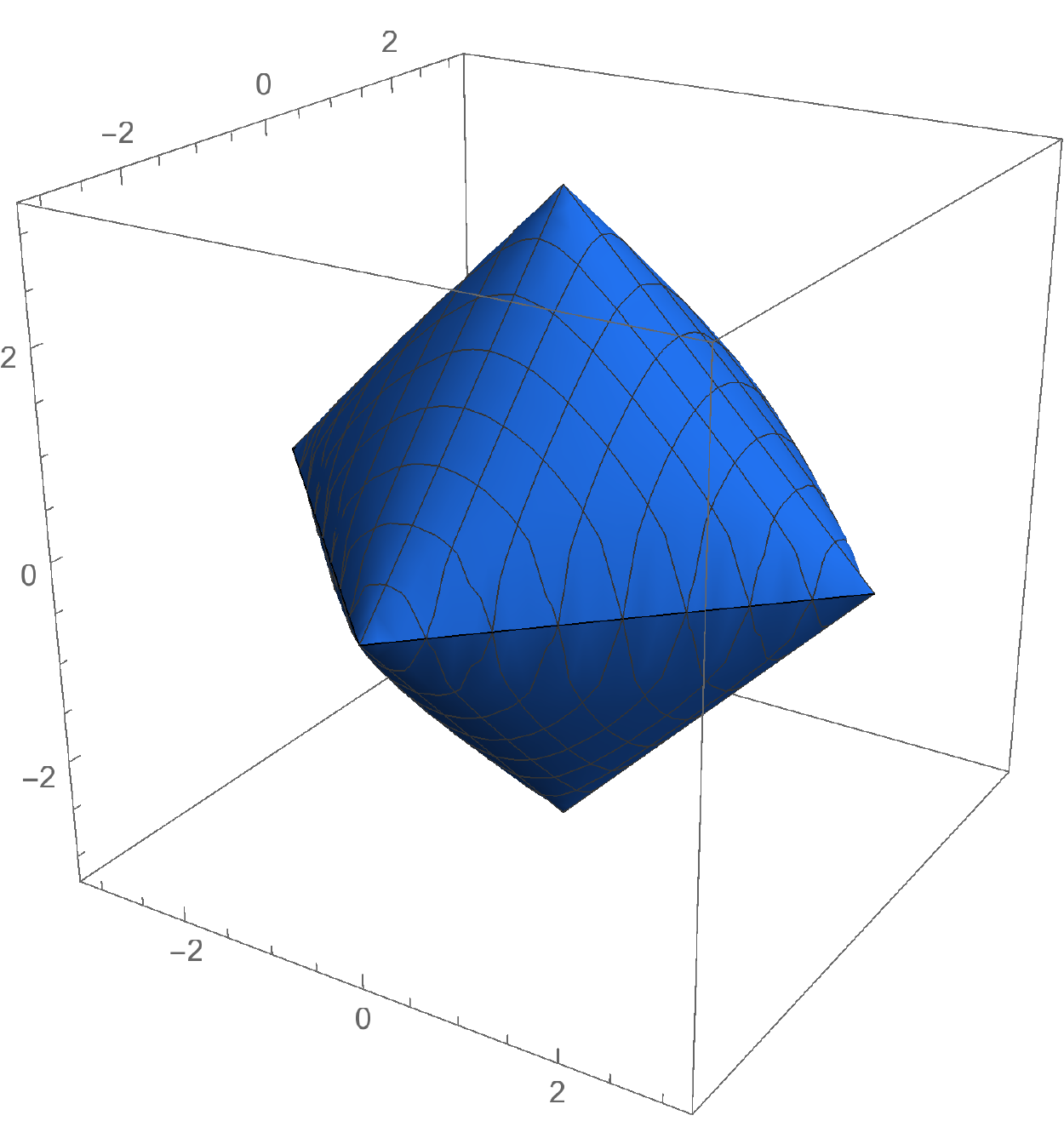}
	\caption{The density of states and the Fermi surfaces of the 3D Hubbard model at \(\mu=0t\) (lower left) and $\mu=-2t$ (lower right plot) .} 
	\label{Hubbard:DOS}
\end{figure}

In order to test the TUFRG in three spatial dimensions, we investigate the isotropic one-band Hubbard model given by the Hamiltonian
\begin{align}\label{Hubbard}
\mathcal{H}= -t\sum_{\langle i,j\rangle,\sigma} c_{i,\sigma}^\dag c_{j,\sigma} + U\sum_{i}n_{i\uparrow}n_{i,\downarrow}
\end{align}
on a simple-cubic three-dimensional lattice. Here \(t\) denotes the hopping amplitude between nearest neighbors and \(c_{i,\sigma}^\dag\,(c_{i,\sigma})\) creates (annihilates) an electron with spin \( \sigma\) on site \(i\). \(U\) is the onsite Coulomb interaction and \(n_{i,\sigma}=c_{i,\sigma}^\dag c_{i,\sigma}\) is the particle density on site \(i\). 
The non-interacting dispersion of the Hamiltonian given in \eqref{Hubbard} is 
\begin{align}
\varepsilon_{\vec{k}}=-2t\left(\cos(k_x)+\cos(k_y)+\cos(k_z)\right) \, .
\end{align} 
The corresponding density of states and Fermi-surface for half filling and for a smaller filling are shown in fig.\ \ref{Hubbard:DOS}. We will use \(t\) as energy unit from now on.\\
Although eq.\ \ref{Hubbard} may be one of the simplest possible interacting many-fermion Hamiltonians, it can not be solved analytically. To obtain the leading ordering tendency and its energy scale we use the TUFRG presented in section \ref{sec:Method}, given by eqns.\ \eqref{TUFRG_eq}, \eqref{bubble_proj} and \eqref{ctop_proj} (with the other channels treated analogously).  
We start the flow equations at a high energy cutoff of \(\Lambda=30\) with an initial interaction \(V_0=U\) for all momentum combinations according to the onsite Hubbard interaction. To obtain the pp- and ph-bubble terms \(\chi\) we neglect self-energy effects, i.e.\ the internal lines are bare Green's functions \(G^\Lambda=G^\Lambda_0\) and single scale propagators \(S^\Lambda=\frac{\mathrm{d}}{\mathrm{d}\Lambda}G^\Lambda_0\). \\
In the FRG flow, we use a smooth frequency cutoff function\cite{Husemann2009} \( C_\Lambda ( \omega_n ) = \frac{\Lambda^2}{\Lambda^2+\omega_n^2}\) on the free Green's function \(G_0\) whose derivative, therefore, is in the single-scale propagator \(S^\Lambda\).  Using frequency-independent vertices at \(T=0\) the bubble terms in Eqs. \ref{bubble_proj} become (before momentum summation)
\begin{flalign}
\begin{split}
&L^\Lambda_\text{pp/ph}(\vec{k},\vec{k}')=\\
&\frac{\mathrm{d}}{\mathrm{d}\Lambda}\int_{-\infty}^{\infty} \frac{\mathrm{d}\omega}{2 \pi} \frac{\omega^2}{\omega^2+\Lambda^2}\frac{1}{i\omega-\varepsilon_{\vec{k}}}\frac{\omega'^2}{\omega'^2+\Lambda^2}\frac{1}{\mp i\omega'-\varepsilon_{\vec{k}'}}
\end{split}
\end{flalign}
where the frequency integral can be solved analytically.\\
In eqns.\ \eqref{bubble_proj} and \eqref{ctop_proj} we use form-factors \(f_i(\vec{q})\) corresponding to the O\(_\text{h}\) symmetry group of the crystal lattice, representing bonds up to the second nearest neighbors, as shown in table \ref{FFtable}. As these form factors are sums of \(\delta\)-functions in real space, the channel-to-channel projections like the one in eq.\ \eqref{ctop_proj} are performed by transforming the right hand side to real space, resulting in a combinatoric problem of multiplying real space representations of the form factors, but without the need to perform a two integrations over momentum space. \\
\begin{table}
	\begin{tabular}{c|c}
		Name & mom. space representation\\ \hline
		\(s\) & \(\textit{const}\)\\
		ext. \(s\) & \(\cos(k_x)+\cos(k_y)+\cos(k_z)\)\\
		\(d_{x^2-y^2}\) & \(\cos(k_x)-\cos(k_y)\) \\
		\(d_{z^2}\) & \(-\cos(k_x)-\cos(k_y)+2\cos(k_z)\)\\
		\(p_x\) & \(\sin(k_x)\) \\
		\(p_y\) & \(\sin(k_y)\) \\
		\(p_z\) & \(\sin(k_z)\) \\
		\(s_3\) & \(\cos(k_x)\cos(k_y)+\cos(k_x)\cos(k_z)+\cos(k_y)\cos(k_z)\)\\
		\(d_{xy}\) & \(\sin(k_x)\sin(k_y)\)\\
		\(d_{xz}\) & \(\sin(k_x)\sin(k_z)\)\\
		\(d_{yz}\) & \(\sin(k_y)\sin(k_z)\)
	\end{tabular}
	\caption{List of form factors $f_i (\vec{k} ) $ used for the calculations, transforming according to irreducible representations of the cubic group $O_h$. For the phase diagram calculation only the first seven corresponding to onsite and nearest neighbor were used, while for the doped system the full list including some second nearest neighbor form-factors was used.}
	\label{FFtable}
\end{table}
For the calculations in section \ref{res:phasediag} a regular momentum grid of \(16^3\) was used for the representation of the vertex, i.e.\ for the first entries in \(\Phi^{P/C/D}\), while the bubbles as in eqns.\ \eqref{bubble_proj} were evaluated on a \(560^3\) momentum mesh which we observed to give well converged results over the full investigated range of \(U\). As an anitferromagnetic state is expected, we only considered the on-site and nearest-neighbor form-factors (the first  in table \ref{FFtable}). The Hubbard interaction \(U\) chosen as \(\Lambda_0=30t\) serves as initial interaction, well above the band width. The flow was stopped when the largest value of a propagator exceeded the maximum of \(V_\text{max}=50t\). The leading component in 
\(\Phi^{P/C/D}\) is used to identify the type of emergent order that is signaled by this flow to strong coupling. The corresponding FRG scale is referred to as critical scale \(\Lambda_c\). In BCS or spin-density-wave meanfield theory, in the weak coupling limit at constant density of states, the band gap equals the critical temperature up to a prefactor of order unity. In analogy to this observation, we expect the critical scale \(\Lambda_c\) to correspond to a critical temperature up to a factor of unity, and, therefore, to be comparable with literature values for critical temperatures.

\section{Results}\label{res:results}

\subsection{Results at half filling}\label{res:phasediag}
\begin{figure}
 \includegraphics[width=0.9\columnwidth]{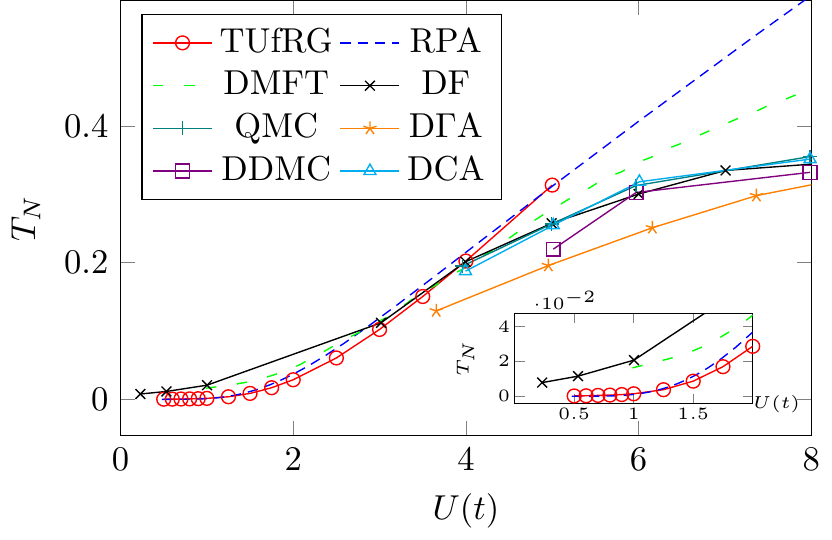}
 \caption{Phase diagram of the three-dimensional Hubbard model. Presented is the critical scale \(\Lambda_c\) for the TUFRG calculation in comparison to the critical temperature for previous calculations. For low interactions the ordering energy scale by obtained by FRG compares well with that obtained by other methods. }
 \label{Phasediag}
\end{figure}

First, we run the 3D-TUFRG for the half-filled, fully nested Fermi surface for varying initial \(U\). This results in a flow to strong coupling with leading AF-SDW correlations at a critical scale \(\Lambda_c (U)\). This critical scale is shown in fig.\ \ref{Phasediag} in comparison to the critical temperatures obtained by RPA\cite{Scalettar1989}, Dynamical Mean Field Theory (DMFT)\cite{Hirschmeier2015}, Dynamical Cluster Approximation (DCA)\cite{Kent2005}, Quantum Monte Carlo (QMC)\cite{Staudt2000}, Determinantal Diagrammatic Monte Carlo (DDMC)\cite{Kozik2013}, Dynamical Vertex Approximation (D\(\Gamma\)A)\cite{Rohringer2011} and Dual Fermions (DF)\cite{Hirschmeier2015}. Note that he RPA values are taken to be the renormalized critical temperatures according to ref. \cite{Freericks1995}, obtained by the Stoner criterion with the bar eparticle-hole bubble and by the dividing the obtained \(T_N\) by three as a cheap way to account for corrections due to the local pairing channel\cite{Kanamori1963}. 
%
Focusing on the shape of the \(\Lambda_c(U)\)-curve, it can be observed that our TUFRG results in \(D=3\) in general compare reasonably well to the other methods at small interactions \(U/t\lesssim 4\) even though the FRG scheme here does not take into account effects of the electronic self-energy and frequency dependence of the effective interactions. Clearly, the present FRG completely misses the plateau in \(T_c\) observed with other methods for \(U/t \approx 8-10\). This failure is expected, as self-energy effects and higher-order vertices are neglected and therefore no pseudogap, Mott transition or local moment formation can be observed. 
Focusing on the quantitative comparison with renormalized RPA and the dual-fermion data at \(U/t\le 4\), we can state the following.
The FRG may have  a lower critical scale than the RPA, because the FRG includes the suppression of the magnetic channel by the full pairing channel, and not only the local contribution\cite{Kanamori1963,Freericks1995}. In addition there may be competing effects from the direct particle-hole channel. The dual-fermion approach of ref.\ \onlinecite{Hirschmeier2015} employs the ladder approximation, i.e.\ is something like RPA with local selfenergy corrections, and quite similar to single-site DMFT at weak \(U\)\cite{Hirschmeier2015}. As the selfenergy effects can be expected to be quite small at these \(U\) it is no surprise that the critical temperatures are closer to the RPA values for intermediate \(U\). We do not have a theory why they DF and DMFT values are significantly higher than the RPA values in the limit \(U\to 0\), but one might speculate that the onsite-repulsion screening in the particle-particle channel, which is captured as well in these approaches on the level of the impurity problem, is for weak \(U\) not as effective as in the renormalized RPA scheme on the extended lattice. 
For the interpretation of the TUFRG ordering scales one should keep in mind that a neglect of dynamic and selfenergy effects is known to lead to slightly higher critical scales for the AF-SDW instability at least for 2D systems\cite{Tagliavini2019}. So our TUFRG curve should be considered as an upper estimate when channel coupling is included.


\begin{figure}
 \includegraphics[width=\columnwidth]{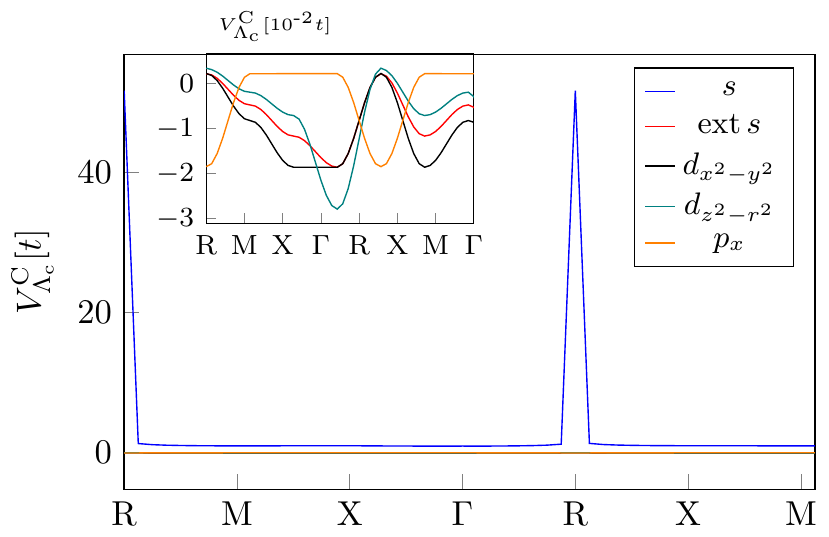}
 \caption{Full vertex projected to the C-channel at the critical scale depending on a main momentum transfer \(\vec{t}\) moving along the high symmetry lines for different form-factors with an initial interaction of \(U=1\,t\).}
 \label{cVertexU1}
\end{figure}
\begin{figure}
 \includegraphics[width=\columnwidth]{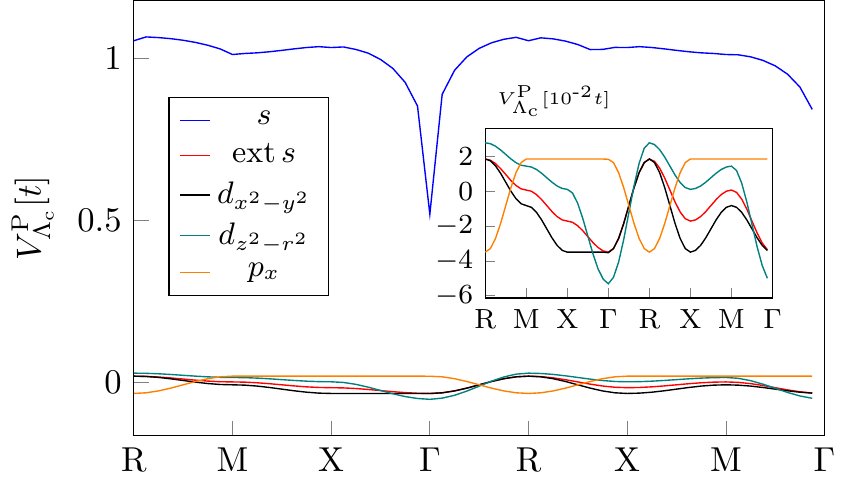}
 \caption{Full vertex projected to the P-channel at the critical scale for a main momentum transfer \(\vec{s}\) along the high symmetry lines for different form-factors with an initial interaction of \(U=1\,t\).}
 \label{pVertexU1}
\end{figure}
Beyond the question of the ordering energy scale, the TUFRG contains information about the relative strength of the individual interaction channels. In fig.\ \ref{cVertexU1} and \ref{pVertexU1} we show the full vertex in the C-(spin-) and P-(pairing-) channel, \(V^\text{C}_{ij}\) and \(V^\text{P}_{ij}\), respectively, for the different form-factors \(i=j\) at the critical scale along the high symmetry axes at \(U=1\). The contributions of combinations \(i\neq j\) are always close to zero and therefore not further considered here. The C-channel exhibits a repulsive divergence for a momentum transfer of \(R=(\pi,\pi,\pi)\) in the (onsite) \(s\)-wave form factor, exceeding the threshold \(V_\text{max}\). On the real-space lattice, the \(s\)-wave form factor means that the spin operators which order and which are build from bilinears in fermion operators are onsite, i.e.\ creation and annihilation operator have the same site index.
The momentum transfer in the C-channel with the peak at $R=(\pi,\pi,\pi )$ is characteristic for antiferromagnetic order, such that a corresponding phase transition is expected to occur at temperatures equal (up to factors of orders of unity) to the energy scale. The contributions of all other form factors to the C-channel, shown  in the inset of Fig. \ref{cVertexU1}, are three orders of magnitudes smaller, varying along the high symmetry lines according to their momentum space shape. The P-channel vertex in fig.\ \ref{pVertexU1} shows a dip of the \(s\)-wave form factor contribution at \(\Gamma\) which still is repulsive. This is the standard suppression of repulsive interactions in this channel. Again, the (onsite) \(s\)-wave form-factor contribution is three orders of magnitude larger than those of the other form factors. 
\begin{figure}
\centering
 \includegraphics[width=0.45\columnwidth]{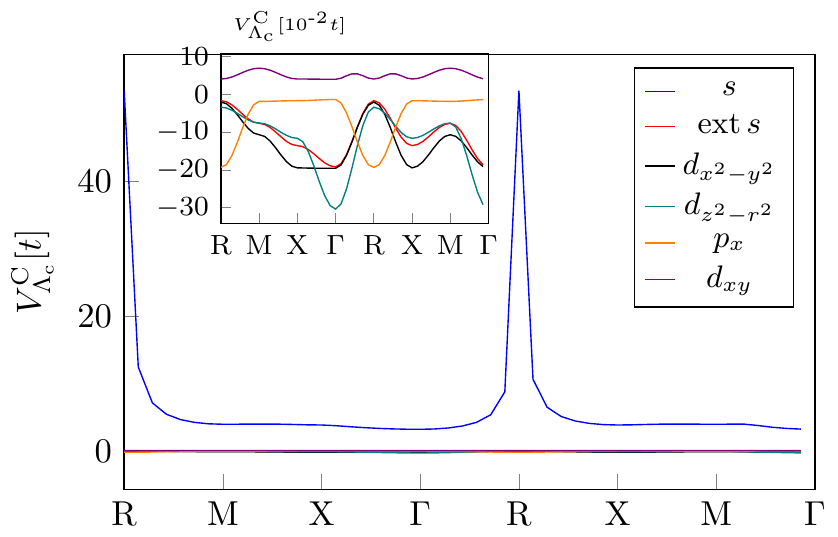}
 \includegraphics[width=0.45\columnwidth]{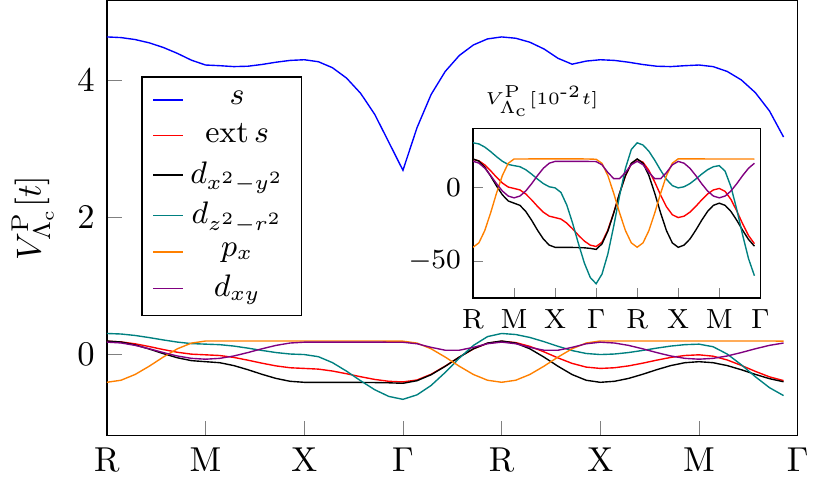}
 \caption{Full vertex projected to the C-channel(left) and P-channel (right) at the critical scale depending on their main momentum transfer moving along the high symmetry lines for different form-factors with an initial interaction of \(U=4\,t\).}
 \label{VertexU4}
\end{figure}
Fig.\ \ref{VertexU4} shows the P- and C-channel at the cutoff scale \(\Lambda_c\) for the system with a higher initial interaction of \(U=4t\). The characteristic features, i.e.\ the peaks in the C-channel at \(R\) and in the P-channel at \(\Gamma\), are the same as in the \(U=1\) case. However, due to the larger initial interaction, there is a broader interaction background of \(U \approx 4t\) in the \(s\)-wave form-factor. This drives the interaction in the other form-factor representations which develop more pronounced features corresponding to its set of form-factors in most cases. 

These results at half filling can be understood as a first sanity check for the 3D-TUFRG scheme. The expected AF-SDW instability is reproduced and quite some details about the wavevector-structure of the effective interaction can be extracted. Rergarding the numbers, the method's quantitative performance seems reasonable in the weak-coupling region for $U\le$ bandwidth/3, even if selfenergy effects and frequency dependences of the vertices are neglected. This is valuable information for more complicated cases, where benchmarks are absent.    

\subsection{Results away from half filling} \label{res:doped}
\begin{figure}
 \includegraphics[width=0.9\columnwidth]{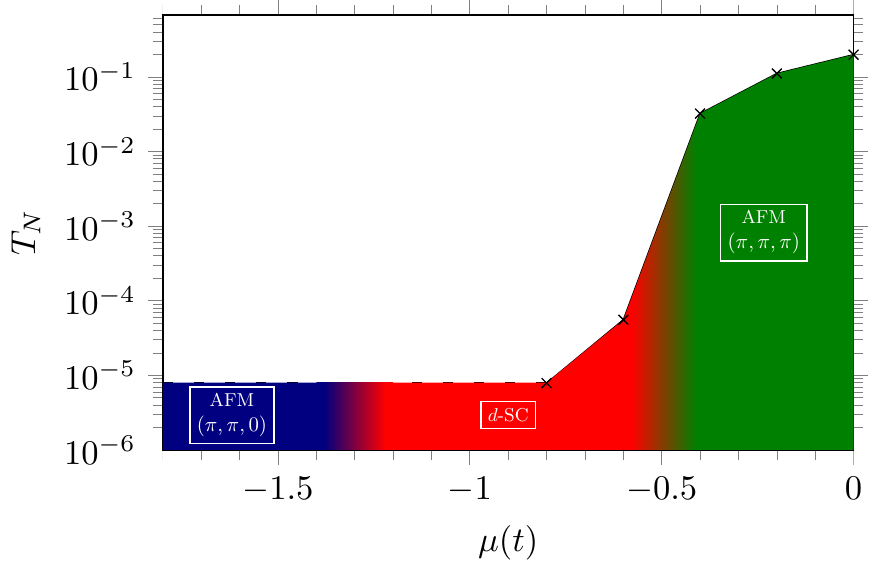}
 \caption{Tentative TUFRG phase diagram of the doped 3D Hubbard model. Close to \(\mu=0.0t\) we observe AFM ordering with \((\pi,\pi,\pi)\)-ordering, around \(\mu=-0.5t\) the ground state becomes \(d_{z^2-r^2}\)-superconducting and around \(\mu=-1.3t\) we observe predominant AFM tendencies, now with \((\pi,\pi,0)\)-ordering (or symmetry-related vectors). Below \(\mu=-1.0t\) the flow does not diverge within a range down to \(\Lambda_c=8*10^{-6}\), hence the plotted line becomes flat. The resulting ordering tendencies are obtained from investigations of the flow and vertex at the smallest \(\Lambda\).}
 \label{doped_phase}
\end{figure}
The occurrence of unconventional pairing in the Hubbard model has been subject to theoretical research already before the advent of the quasi-2D cuprate high-temperature superconductors. In the mid-1980s, Scalapino and collaborators\cite{Scalapino1986} investigated the doped 3D Hubbard model in spin-fluctuation theory and determined the leading pairing instabilities. Besides of the AFM state close to half-filling they detected pairing channels corresponding to nearest neighbor pairs as well as pairing tendencies with some second nearest-neighbor form-factors like \( d_{xy} \sim \sin(k_x)\sin(k_y)\). Hence we perform the calculations away from half filling by using the extended set of form-factors corresponding to on-site, nearest- and next-nearest neighbors in table \ref{FFtable}. For these calculations the vertex was represented by a \(14^3\) momentum grid, while the propagator bubble was evaluated on a \(490^3\) grid. If the system is hole-doped away from half-filling, the critical scale decreases, as shown in fig.\ \ref{doped_phase}. The momentum resolution, limited by computing resources, prohibits statements of critical scales lower than \(\approx 10^{-5}\), such that the flow ended due to this criterion and not by the divergence of the vertex for \(\mu\lesssim -1.0t\). The region, where the flow did not diverge within the observable range is therefore displayed by a dashed line. However, from the observation of the flow in these cases educated guesses on the dominating order are still possible.\\
\begin{figure}
 \includegraphics[width=0.45\columnwidth]{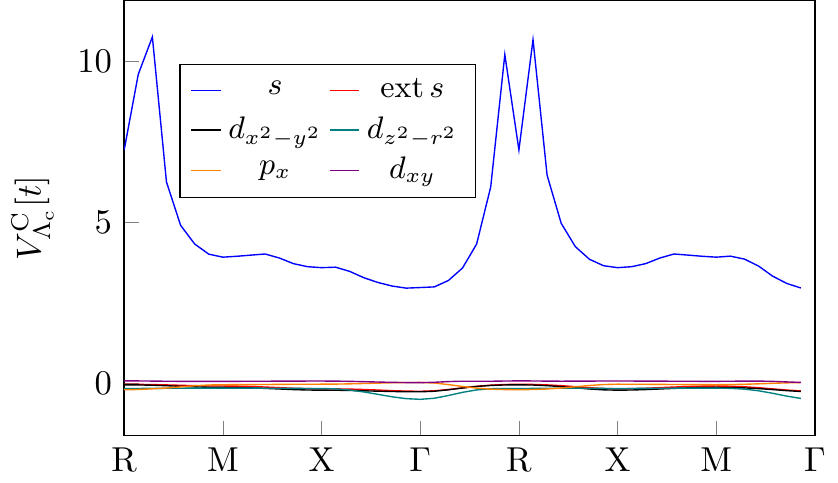}
 \includegraphics[width=0.45\columnwidth]{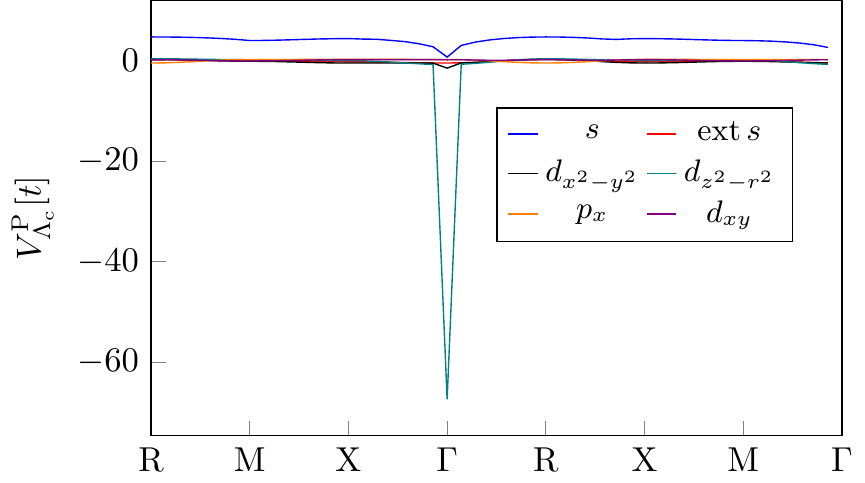}
 \caption{Full TUFRG vertex projected to the C-channel(left) and P-channel (right) at the critical scale depending on their main momentum transfer moving along the high symmetry lines for different form-factors with an initial interaction of \(U=4\,t\) and chemical potential \(\mu=-0.6\,t\).}
 \label{VertexU4mu06}
\end{figure}
\begin{figure}
 \includegraphics[width=0.45\columnwidth]{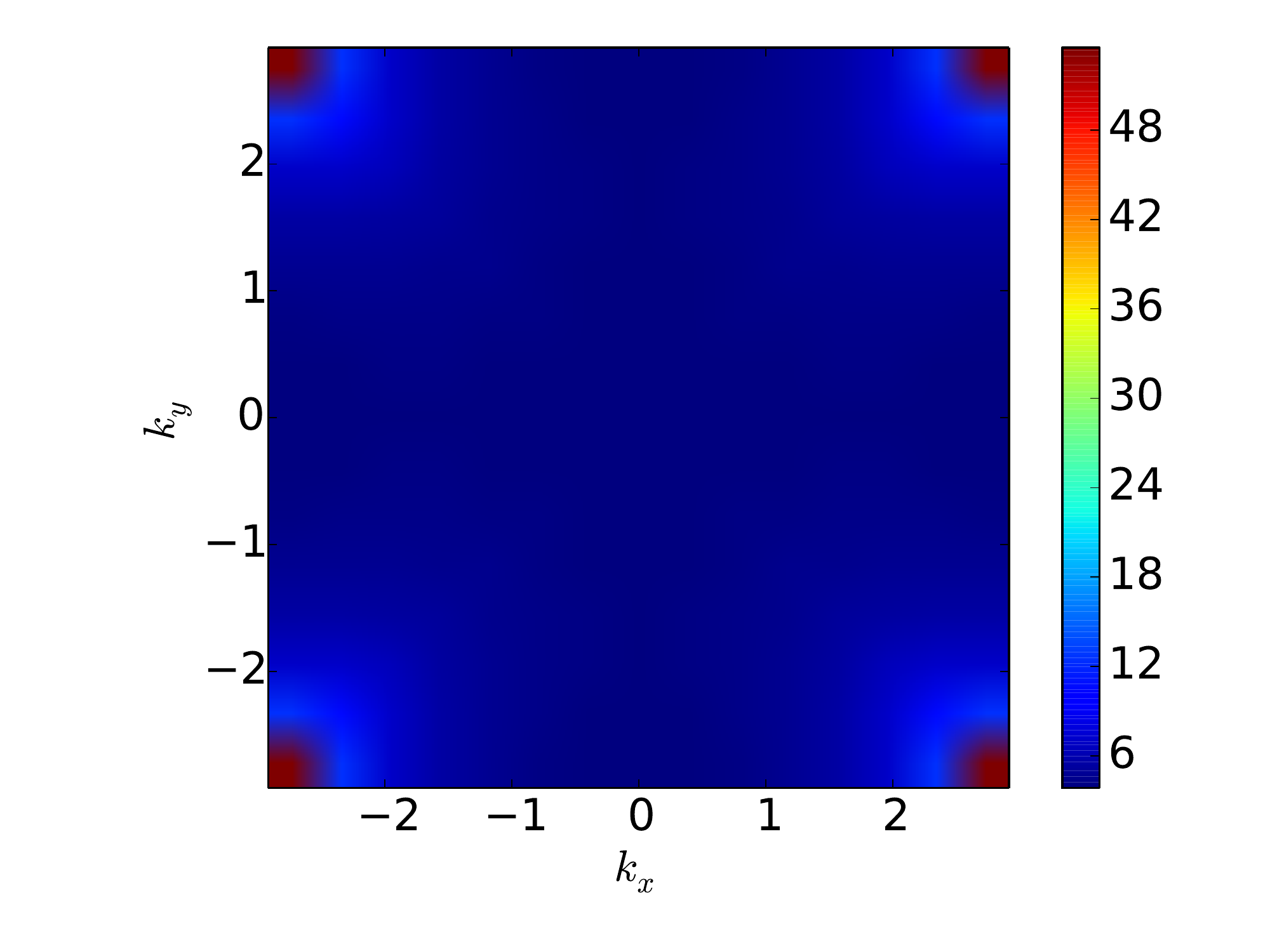}
 \includegraphics[width=0.45\columnwidth]{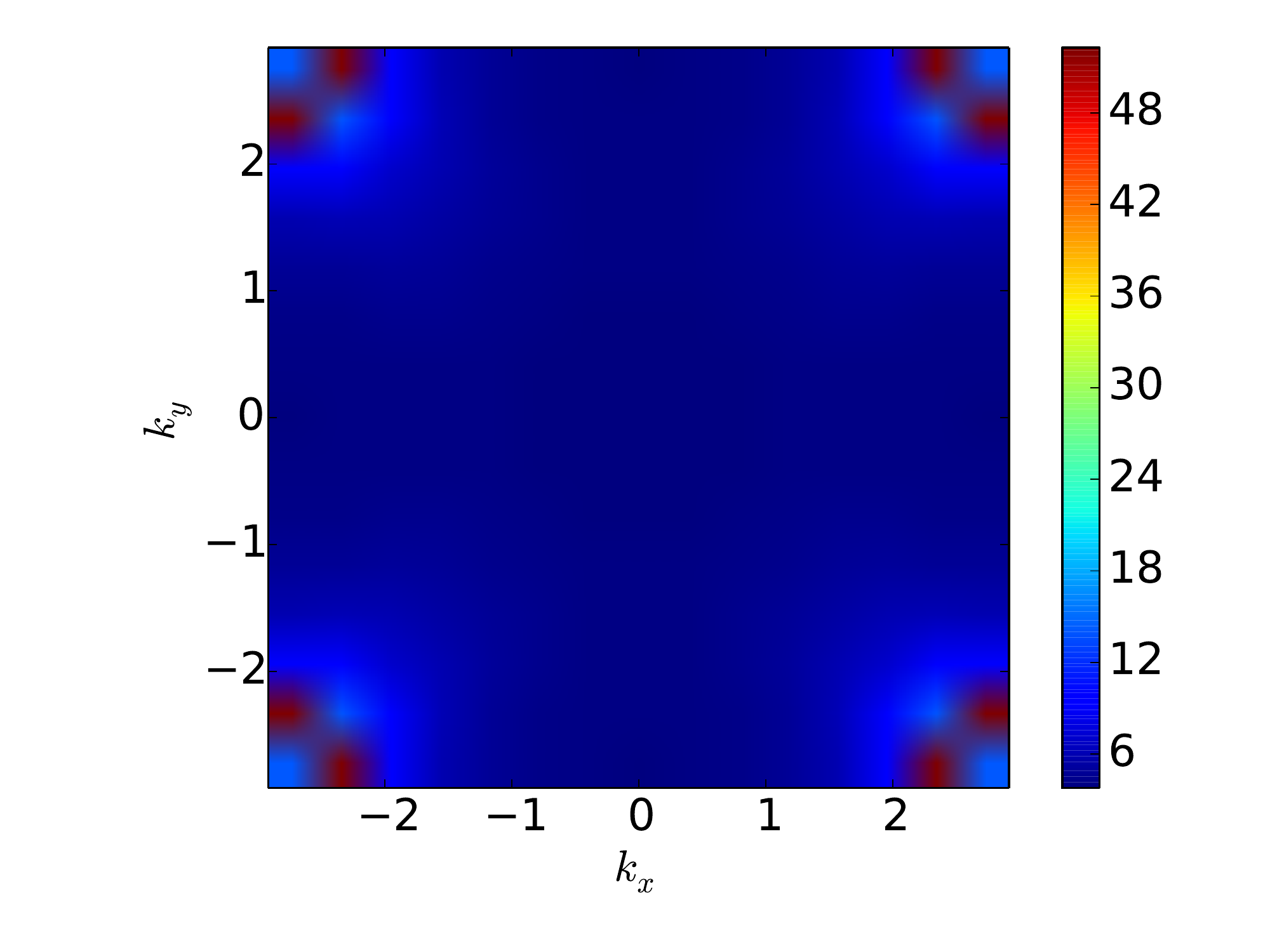}
 \includegraphics[width=0.45\columnwidth]{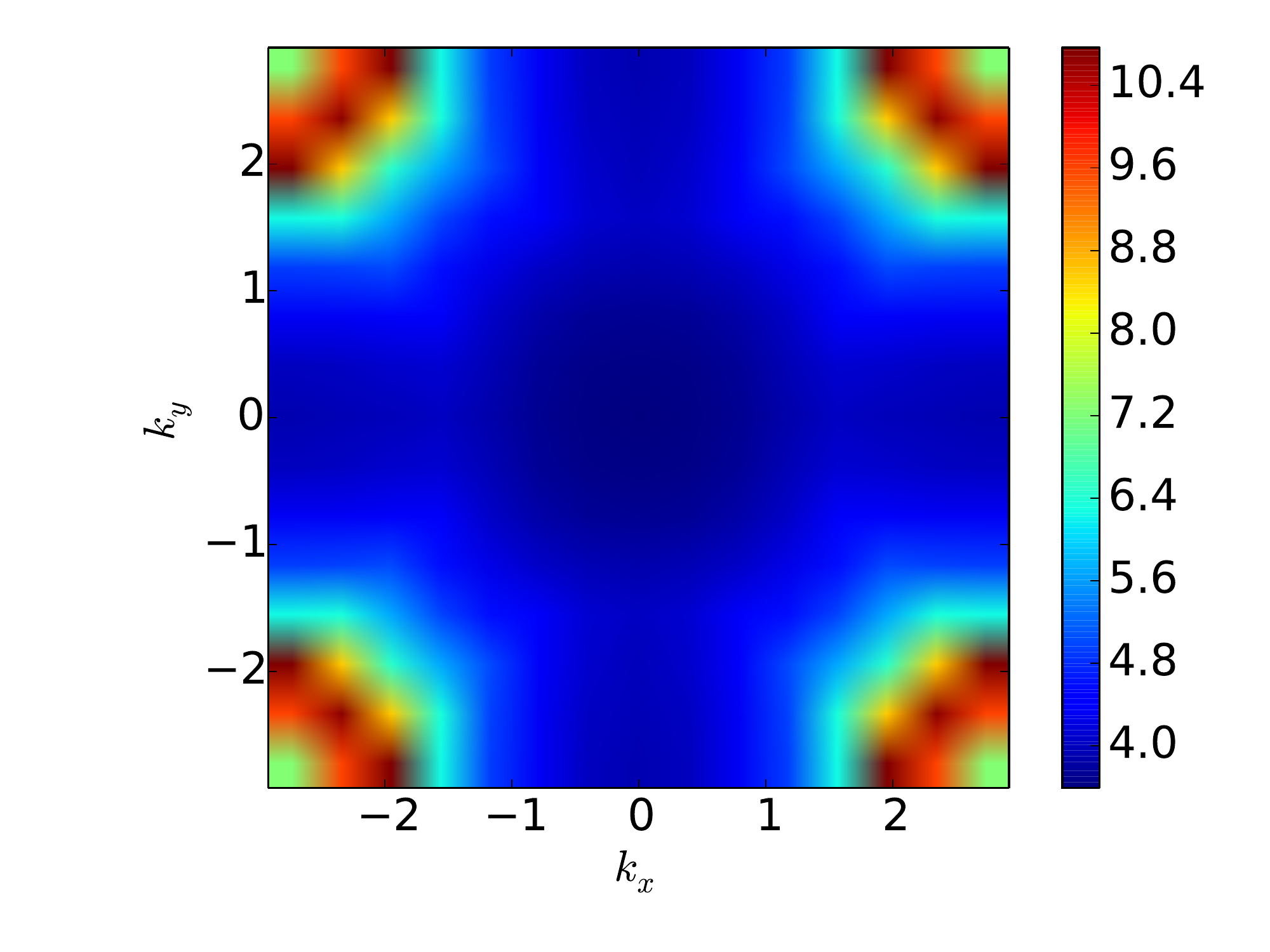}
 \includegraphics[width=0.45\columnwidth]{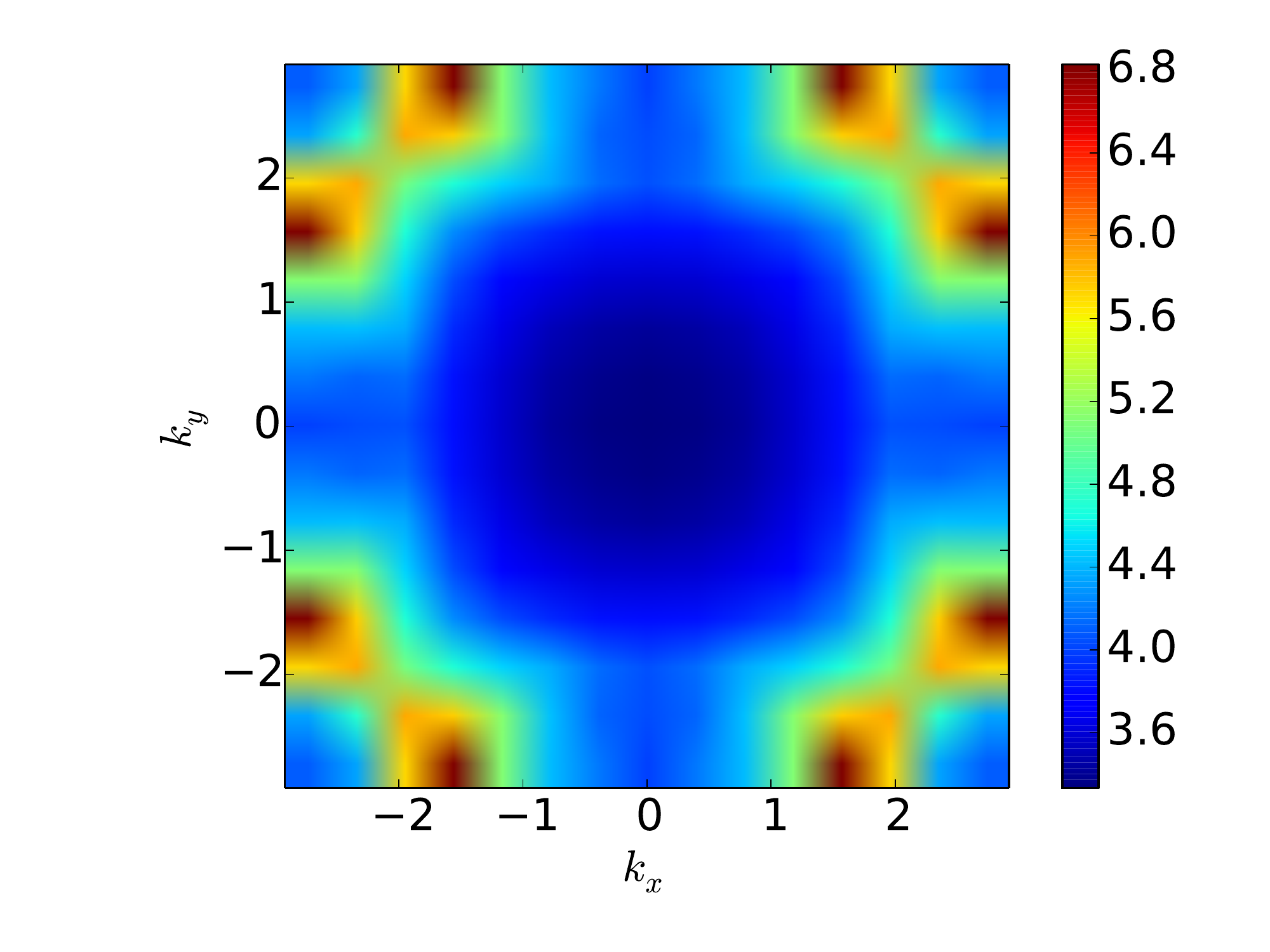}
 \includegraphics[width=0.45\columnwidth]{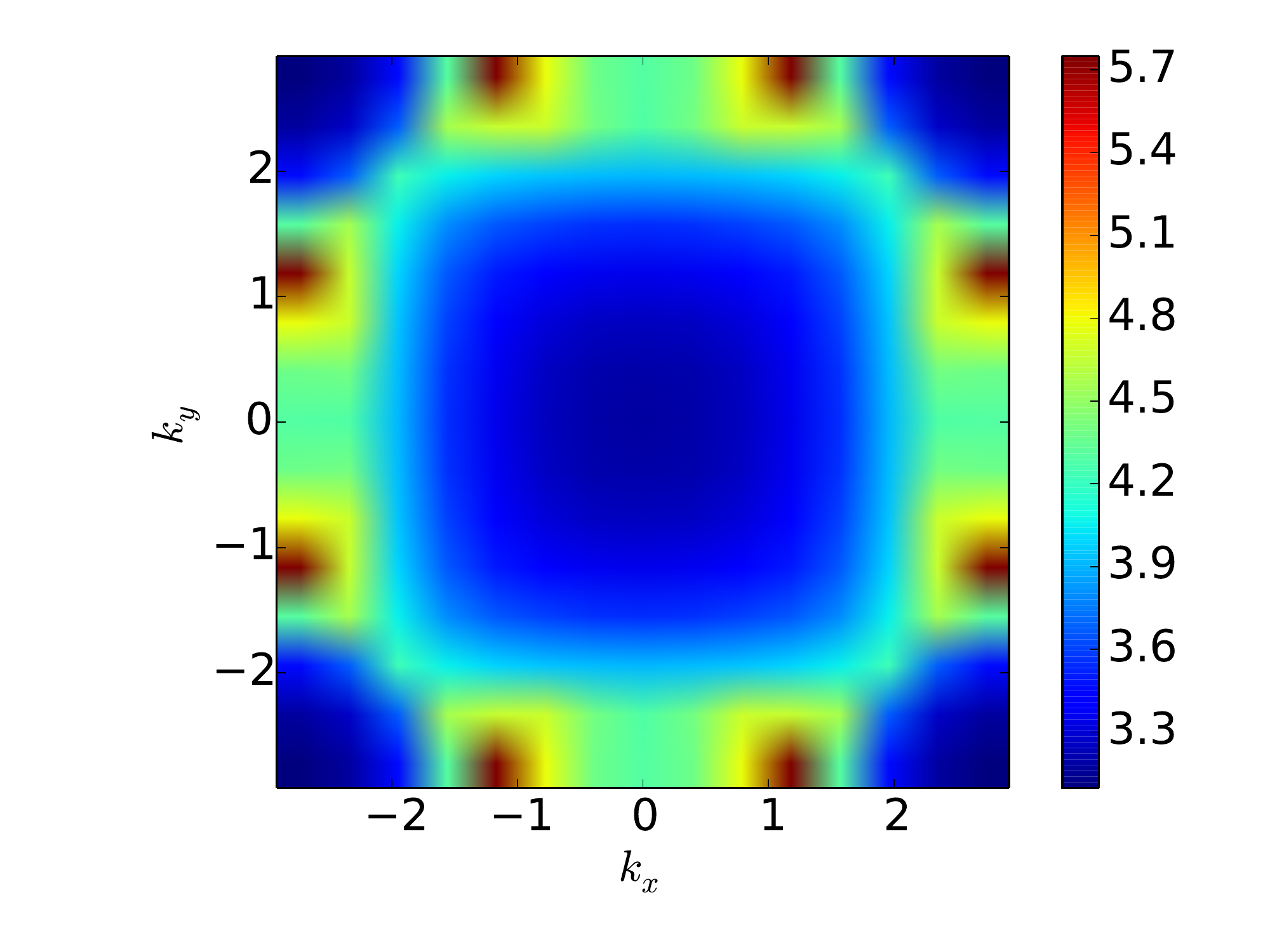}
 \includegraphics[width=0.45\columnwidth]{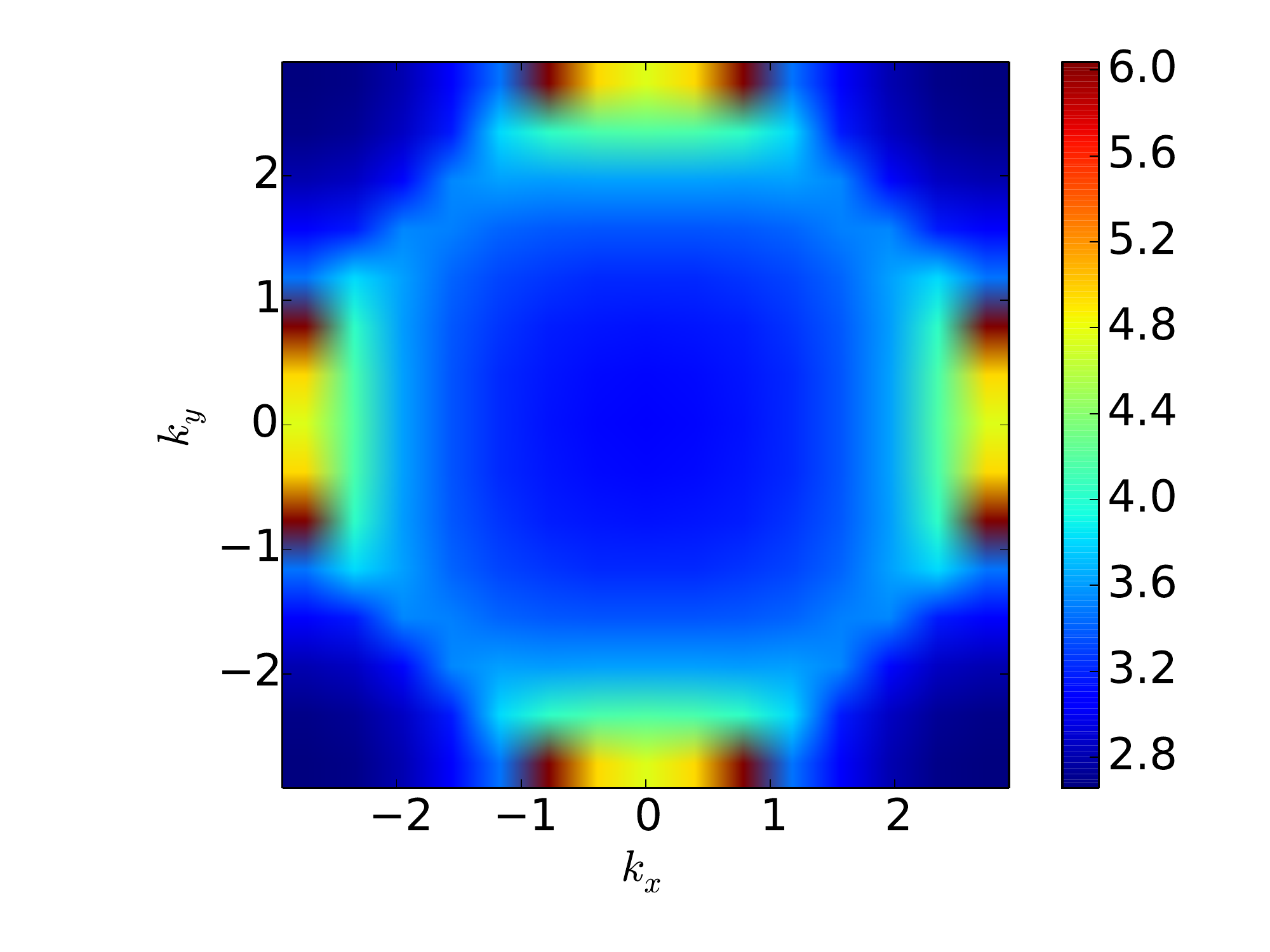}
 \caption{Cut in the \(z=\pi\)-plane of the full vertex in the C-channel with the on-site formfactor at \(U=4t\) and \(\mu=0t,\,-0.4t,\,-0.6t,\,-1.0t,\,-1.4t,\,-1.8t\) from left to right and top to bottom. }
 \label{z=pi-plot}
\end{figure}
For \(\mu\gtrsim-0.5t\) we still observe a transition to an AFM state with a contribution of the vertex similar to those in the preceding section. As the doping moves the Fermi surface away from perfect nesting, the AFM state becomes incommensurate and the critical scale for the run-away flow decreases. In figure \ref{z=pi-plot} (top row) we show a cut along the \(z=\pi\)-plane of the C-channel in the leading on-site form-factor both, for \(\mu=0t\) and \(\mu=-0.4t\). The plots clearly show, that the AFM nesting becomes incommensurate, as the peak splits up and moves slightly along \((x,\pi,\pi)\) and \(\pi,y,\pi)\) towards both \(M\)-points on this plane. This split to different set of nestings then causes the drop in critical scale.\\
For \(-1.0t\lesssim\mu\lesssim-0.5t\) we observe a leading instability in the pairing channel, with a critical scale of the order \(\approx 10^{-5}\). A transition between the AFM and \(d\)-wave state is predicted from diagrammatic calculations\cite{Scalapino1986} to happen at \(\mu=-0.8t\) slightly below our values. Therefore the inter-channel couplings present in our calculations but not in the previous spin-fluctuation theories seem to enlarge the \(d\)-wave regime. In this parameter region the leading instability occurs in the pairing channel with \(d_{z^2-r^2}\) form-factors at \(\Gamma\), indicating a \(d_{z^2-r^2}\)-superconducting phase (see fig.\ \ref{VertexU4mu06}). 
Of course, there are two more symmetry-related form-factors with the same pairing strength that can be obtained from rotating the  
\(d_{z^2-r^2}\)-function to point along the \(x\)- or \(y\)-axis. These will then admix to the \(d_{z^2-r^2}\)-function the other basis function of this 2-dimensional irreducible representation of O\(_h\), customarily chosen as having  \(d_{x^2-y^2}\)-symmetry. In figure \ref{VertexU4mu06} the basic feature of a divergency of the vertex in this form-factor basis can be seen, however with a much smaller pairing strength.
This is presumably because the corresponding gap function would have longer nodal lines on the Fermi surface than the  \(d_{z^2-r^2}\)-function. Regarding the C-channel in this region, there is still a subdominant incommensurate AFM ordering tendency which continues the trend of peaks moving from \(R\) towards \(M\) (see fig.\ \ref{z=pi-plot}).\\


For studying the leading order in the region of \(\mu\lesssim-1.0t\) we considered the change of the propagators \(P\), \(C\) and \(D\) as well as the vertex at \(\Lambda=10^{-5}\) where we ended the flow. For \(\mu=-1.2\) the leading contribution to the flow and the signature in the vertex still correspond to a \(d_{z^2-r^2}\)-SC phase. However, for even smaller \(\mu\) the main contribution to the flow comes from the \(C\)-channel. The vertex for this scale shows a peak close to \(M=(\pi,\pi,0)\), i.e.\ the system tends to an antiferromagnetic order with in-plane ordering vector. Regarding the trend of increased doping for the AFM ordering peak, it moved from \(R\) in the undoped region towards \(M\) for \(\mu\approx-2t\). Close to this chemical potential, the two incommensurate peaks corresponding to e.g.\ \(\pi,\pi,\pm\eta)\), with \(\eta\) a small positive number, get close to each other and thus strengthen the in-plane AFM tendency. This then destroys the \(d\)-SC order which still causes signatures in the vertex. As the Fermi-surface of the 3D Hubbard model becomes an inflated octahedron at \(\mu=-2.0t\) (see Fig.\ref{Hubbard:DOS}) with edges corresponding to a perfectly nested 2D square-lattice Fermi surface, a leading AFM order at the \(M\)-points can be expected. Further investigations might also investigate the situation for even larger doping with \(\mu\) below \(-2.0t\).

\section{Conclusion}\label{conclusion}
We applied the truncated-unity functional renormalization (TUFRG) to the three-dimensional Hubbard model. This study serves as a first demonstration that this method allows us to investigate three-dimensional lattice-fermion models with functional renormalization group methods. Here, we basically aimed to show that the scheme produces the correct physics and that it may even be used on the quantitative level. 

At half band filling and perfect nesting we observed the well-known antiferromagnetic ground state of the 3D Hubbard model  by analyzing the flow to strong coupling in the TUFRG. By using the TUFRG critical scale as an estimate for the N\'eel ordering temperature \(T_\text{N}\) we could reasonably well reproduce the behavior of the \(T_\text{N}(U)\) computed by other numerical techniques for weak coupling strengths \(U\lesssim 4t\), although we used a set of form-factors limited to nearest neighbors. However, as expected, due to the approximations involved, our approach does not reproduce the plateau of \(T_\text{N}\) at intermediate coupling strength and the crossover to a decreasing ordering scale \(\sim 1/U\) at larger \(U\). Electron- or hole-doping of the system leads to a significant decrease of the critical scale as the Fermi surface does not provide perfect nesting anymore. For \(\mu<-0.5\) the ground state becomes \(d_{z^2-r^2}\)-wave superconducting in good agreement to the predictions of reference \onlinecite{Scalapino1986}. For a doping towards \(\mu\approx-2t\) this \(d\)-wave superconducting order becomes suppressed by an in-plane antiferromagnetic ordering. Further investigations of the doped 3D Hubbard model reducing the approximations are under way.

Due to the numerical advantages of the TUFRG approach, the inclusion of selfenergy feedback, frequency dependence and possibly multi-loop effects\cite{Tagliavini2019} is conceivable. This will further increase the quantitative control of the method. It may also be interesting to analyze in more detail the critical behavior displayed by the TUFRG and to understand to which extent non-field-behavior\cite{Rohringer2011} can be reproduced depending on the approximation level. 
Furthermore, extensions to models with several orbitals should be feasible, if necessary with some compromises in the momentum resolution.

\section{Acknowledgements}
The authors gratefully acknowledge financial support through the DFG research training group 1995, DFG HO-2422/12-1 and the computing time granted through JARA-HPC on the supercomputer JURECA at Forschungszentrum Jülich\cite{jureca}. We thank D.\ Rohe, S.\ Andergassen, S.\ Blügel, C.\ Eckhardt, C.\ Hille, G.\ Schober and A.\ Tagliavini for discussions.

\bibliography{3D_lit}

\begin{thebibliography}{33}%
\makeatletter
\providecommand \@ifxundefined [1]{%
 \@ifx{#1\undefined}
}%
\providecommand \@ifnum [1]{%
 \ifnum #1\expandafter \@firstoftwo
 \else \expandafter \@secondoftwo
 \fi
}%
\providecommand \@ifx [1]{%
 \ifx #1\expandafter \@firstoftwo
 \else \expandafter \@secondoftwo
 \fi
}%
\providecommand \natexlab [1]{#1}%
\providecommand \enquote  [1]{``#1''}%
\providecommand \bibnamefont  [1]{#1}%
\providecommand \bibfnamefont [1]{#1}%
\providecommand \citenamefont [1]{#1}%
\providecommand \href@noop [0]{\@secondoftwo}%
\providecommand \href [0]{\begingroup \@sanitize@url \@href}%
\providecommand \@href[1]{\@@startlink{#1}\@@href}%
\providecommand \@@href[1]{\endgroup#1\@@endlink}%
\providecommand \@sanitize@url [0]{\catcode `\\12\catcode `\$12\catcode
  `\&12\catcode `\#12\catcode `\^12\catcode `\_12\catcode `\%12\relax}%
\providecommand \@@startlink[1]{}%
\providecommand \@@endlink[0]{}%
\providecommand \url  [0]{\begingroup\@sanitize@url \@url }%
\providecommand \@url [1]{\endgroup\@href {#1}{\urlprefix }}%
\providecommand \urlprefix  [0]{URL }%
\providecommand \Eprint [0]{\href }%
\providecommand \doibase [0]{http://dx.doi.org/}%
\providecommand \selectlanguage [0]{\@gobble}%
\providecommand \bibinfo  [0]{\@secondoftwo}%
\providecommand \bibfield  [0]{\@secondoftwo}%
\providecommand \translation [1]{[#1]}%
\providecommand \BibitemOpen [0]{}%
\providecommand \bibitemStop [0]{}%
\providecommand \bibitemNoStop [0]{.\EOS\space}%
\providecommand \EOS [0]{\spacefactor3000\relax}%
\providecommand \BibitemShut  [1]{\csname bibitem#1\endcsname}%
\let\auto@bib@innerbib\@empty
\bibitem [{\citenamefont {Georges}\ and\ \citenamefont
  {Kotliar}(1992)}]{Georges1992}%
  \BibitemOpen
  \bibfield  {author} {\bibinfo {author} {\bibfnamefont {A.}~\bibnamefont
  {Georges}}\ and\ \bibinfo {author} {\bibfnamefont {G.}~\bibnamefont
  {Kotliar}},\ }\href {\doibase 10.1103/PhysRevB.45.6479} {\bibfield  {journal}
  {\bibinfo  {journal} {Phys. Rev. B}\ }\textbf {\bibinfo {volume} {45}},\
  \bibinfo {pages} {6479} (\bibinfo {year} {1992})}\BibitemShut {NoStop}%
\bibitem [{\citenamefont {Hirsch}(1987)}]{Hirsch1987}%
  \BibitemOpen
  \bibfield  {author} {\bibinfo {author} {\bibfnamefont {J.~E.}\ \bibnamefont
  {Hirsch}},\ }\href {\doibase 10.1103/PhysRevB.35.1851} {\bibfield  {journal}
  {\bibinfo  {journal} {Phys. Rev. B}\ }\textbf {\bibinfo {volume} {35}},\
  \bibinfo {pages} {1851} (\bibinfo {year} {1987})}\BibitemShut {NoStop}%
\bibitem [{\citenamefont {Mart\'{\i}n-Rodero}\ and\ \citenamefont
  {Flores}(1992)}]{Rodero1992}%
  \BibitemOpen
  \bibfield  {author} {\bibinfo {author} {\bibfnamefont {A.}~\bibnamefont
  {Mart\'{\i}n-Rodero}}\ and\ \bibinfo {author} {\bibfnamefont
  {F.}~\bibnamefont {Flores}},\ }\href {\doibase 10.1103/PhysRevB.45.13008}
  {\bibfield  {journal} {\bibinfo  {journal} {Phys. Rev. B}\ }\textbf {\bibinfo
  {volume} {45}},\ \bibinfo {pages} {13008} (\bibinfo {year}
  {1992})}\BibitemShut {NoStop}%
\bibitem [{\citenamefont {Freericks}\ and\ \citenamefont
  {Jarrell}(1995)}]{Freericks1995}%
  \BibitemOpen
  \bibfield  {author} {\bibinfo {author} {\bibfnamefont {J.~K.}\ \bibnamefont
  {Freericks}}\ and\ \bibinfo {author} {\bibfnamefont {M.}~\bibnamefont
  {Jarrell}},\ }\href {\doibase 10.1103/PhysRevLett.74.186} {\bibfield
  {journal} {\bibinfo  {journal} {Phys. Rev. Lett.}\ }\textbf {\bibinfo
  {volume} {74}},\ \bibinfo {pages} {186} (\bibinfo {year} {1995})}\BibitemShut
  {NoStop}%
\bibitem [{\citenamefont {Sandvik}(1998)}]{Sandvik1998}%
  \BibitemOpen
  \bibfield  {author} {\bibinfo {author} {\bibfnamefont {A.~W.}\ \bibnamefont
  {Sandvik}},\ }\href {\doibase 10.1103/PhysRevLett.80.5196} {\bibfield
  {journal} {\bibinfo  {journal} {Phys. Rev. Lett.}\ }\textbf {\bibinfo
  {volume} {80}},\ \bibinfo {pages} {5196} (\bibinfo {year}
  {1998})}\BibitemShut {NoStop}%
\bibitem [{\citenamefont {Affleck}\ \emph {et~al.}(1988)\citenamefont
  {Affleck}, \citenamefont {Zou}, \citenamefont {Hsu},\ and\ \citenamefont
  {Anderson}}]{Affleck1988}%
  \BibitemOpen
  \bibfield  {author} {\bibinfo {author} {\bibfnamefont {I.}~\bibnamefont
  {Affleck}}, \bibinfo {author} {\bibfnamefont {Z.}~\bibnamefont {Zou}},
  \bibinfo {author} {\bibfnamefont {T.}~\bibnamefont {Hsu}}, \ and\ \bibinfo
  {author} {\bibfnamefont {P.~W.}\ \bibnamefont {Anderson}},\ }\href {\doibase
  10.1103/PhysRevB.38.745} {\bibfield  {journal} {\bibinfo  {journal} {Phys.
  Rev. B}\ }\textbf {\bibinfo {volume} {38}},\ \bibinfo {pages} {745} (\bibinfo
  {year} {1988})}\BibitemShut {NoStop}%
\bibitem [{\citenamefont {{Staudt, R.}}\ \emph {et~al.}(2000)\citenamefont
  {{Staudt, R.}}, \citenamefont {{Dzierzawa, M.}},\ and\ \citenamefont
  {{Muramatsu, A.}}}]{Staudt2000}%
  \BibitemOpen
  \bibfield  {author} {\bibinfo {author} {\bibnamefont {{Staudt, R.}}},
  \bibinfo {author} {\bibnamefont {{Dzierzawa, M.}}}, \ and\ \bibinfo {author}
  {\bibnamefont {{Muramatsu, A.}}},\ }\href {\doibase 10.1007/s100510070120}
  {\bibfield  {journal} {\bibinfo  {journal} {Eur. Phys. J. B}\ }\textbf
  {\bibinfo {volume} {17}},\ \bibinfo {pages} {411} (\bibinfo {year}
  {2000})}\BibitemShut {NoStop}%
\bibitem [{\citenamefont {Fuchs}\ \emph {et~al.}(2011)\citenamefont {Fuchs},
  \citenamefont {Gull}, \citenamefont {Troyer}, \citenamefont {Jarrell},\ and\
  \citenamefont {Pruschke}}]{Fuchs2011}%
  \BibitemOpen
  \bibfield  {author} {\bibinfo {author} {\bibfnamefont {S.}~\bibnamefont
  {Fuchs}}, \bibinfo {author} {\bibfnamefont {E.}~\bibnamefont {Gull}},
  \bibinfo {author} {\bibfnamefont {M.}~\bibnamefont {Troyer}}, \bibinfo
  {author} {\bibfnamefont {M.}~\bibnamefont {Jarrell}}, \ and\ \bibinfo
  {author} {\bibfnamefont {T.}~\bibnamefont {Pruschke}},\ }\href {\doibase
  10.1103/PhysRevB.83.235113} {\bibfield  {journal} {\bibinfo  {journal} {Phys.
  Rev. B}\ }\textbf {\bibinfo {volume} {83}},\ \bibinfo {pages} {235113}
  (\bibinfo {year} {2011})}\BibitemShut {NoStop}%
\bibitem [{\citenamefont {Hirschmeier}\ \emph {et~al.}(2015)\citenamefont
  {Hirschmeier}, \citenamefont {Hafermann}, \citenamefont {Gull}, \citenamefont
  {Lichtenstein},\ and\ \citenamefont {Antipov}}]{Hirschmeier2015}%
  \BibitemOpen
  \bibfield  {author} {\bibinfo {author} {\bibfnamefont {D.}~\bibnamefont
  {Hirschmeier}}, \bibinfo {author} {\bibfnamefont {H.}~\bibnamefont
  {Hafermann}}, \bibinfo {author} {\bibfnamefont {E.}~\bibnamefont {Gull}},
  \bibinfo {author} {\bibfnamefont {A.~I.}\ \bibnamefont {Lichtenstein}}, \
  and\ \bibinfo {author} {\bibfnamefont {A.~E.}\ \bibnamefont {Antipov}},\
  }\href {\doibase 10.1103/PhysRevB.92.144409} {\bibfield  {journal} {\bibinfo
  {journal} {Phys. Rev. B}\ }\textbf {\bibinfo {volume} {92}},\ \bibinfo
  {pages} {144409} (\bibinfo {year} {2015})}\BibitemShut {NoStop}%
\bibitem [{\citenamefont {Kent}\ \emph {et~al.}(2005)\citenamefont {Kent},
  \citenamefont {Jarrell}, \citenamefont {Maier},\ and\ \citenamefont
  {Pruschke}}]{Kent2005}%
  \BibitemOpen
  \bibfield  {author} {\bibinfo {author} {\bibfnamefont {P.~R.~C.}\
  \bibnamefont {Kent}}, \bibinfo {author} {\bibfnamefont {M.}~\bibnamefont
  {Jarrell}}, \bibinfo {author} {\bibfnamefont {T.~A.}\ \bibnamefont {Maier}},
  \ and\ \bibinfo {author} {\bibfnamefont {T.}~\bibnamefont {Pruschke}},\
  }\href {\doibase 10.1103/PhysRevB.72.060411} {\bibfield  {journal} {\bibinfo
  {journal} {Phys. Rev. B}\ }\textbf {\bibinfo {volume} {72}},\ \bibinfo
  {pages} {060411} (\bibinfo {year} {2005})}\BibitemShut {NoStop}%
\bibitem [{\citenamefont {Kozik}\ \emph {et~al.}(2013)\citenamefont {Kozik},
  \citenamefont {Burovski}, \citenamefont {Scarola},\ and\ \citenamefont
  {Troyer}}]{Kozik2013}%
  \BibitemOpen
  \bibfield  {author} {\bibinfo {author} {\bibfnamefont {E.}~\bibnamefont
  {Kozik}}, \bibinfo {author} {\bibfnamefont {E.}~\bibnamefont {Burovski}},
  \bibinfo {author} {\bibfnamefont {V.~W.}\ \bibnamefont {Scarola}}, \ and\
  \bibinfo {author} {\bibfnamefont {M.}~\bibnamefont {Troyer}},\ }\href
  {\doibase 10.1103/PhysRevB.87.205102} {\bibfield  {journal} {\bibinfo
  {journal} {Phys. Rev. B}\ }\textbf {\bibinfo {volume} {87}},\ \bibinfo
  {pages} {205102} (\bibinfo {year} {2013})}\BibitemShut {NoStop}%
\bibitem [{\citenamefont {Rohringer}\ \emph {et~al.}(2011)\citenamefont
  {Rohringer}, \citenamefont {Toschi}, \citenamefont {Katanin},\ and\
  \citenamefont {Held}}]{Rohringer2011}%
  \BibitemOpen
  \bibfield  {author} {\bibinfo {author} {\bibfnamefont {G.}~\bibnamefont
  {Rohringer}}, \bibinfo {author} {\bibfnamefont {A.}~\bibnamefont {Toschi}},
  \bibinfo {author} {\bibfnamefont {A.}~\bibnamefont {Katanin}}, \ and\
  \bibinfo {author} {\bibfnamefont {K.}~\bibnamefont {Held}},\ }\href {\doibase
  10.1103/PhysRevLett.107.256402} {\bibfield  {journal} {\bibinfo  {journal}
  {Phys. Rev. Lett.}\ }\textbf {\bibinfo {volume} {107}},\ \bibinfo {pages}
  {256402} (\bibinfo {year} {2011})}\BibitemShut {NoStop}%
\bibitem [{\citenamefont {van Dongen}(1991)}]{Dongen1991}%
  \BibitemOpen
  \bibfield  {author} {\bibinfo {author} {\bibfnamefont {P.~G.~J.}\
  \bibnamefont {van Dongen}},\ }\href {\doibase 10.1103/PhysRevLett.67.757}
  {\bibfield  {journal} {\bibinfo  {journal} {Phys. Rev. Lett.}\ }\textbf
  {\bibinfo {volume} {67}},\ \bibinfo {pages} {757} (\bibinfo {year}
  {1991})}\BibitemShut {NoStop}%
\bibitem [{\citenamefont {van Dongen}(1994)}]{Dongen1994}%
  \BibitemOpen
  \bibfield  {author} {\bibinfo {author} {\bibfnamefont {P.~G.~J.}\
  \bibnamefont {van Dongen}},\ }\href {\doibase 10.1103/PhysRevB.50.14016}
  {\bibfield  {journal} {\bibinfo  {journal} {Phys. Rev. B}\ }\textbf {\bibinfo
  {volume} {50}},\ \bibinfo {pages} {14016} (\bibinfo {year}
  {1994})}\BibitemShut {NoStop}%
\bibitem [{\citenamefont {Scalapino}\ \emph {et~al.}(1986)\citenamefont
  {Scalapino}, \citenamefont {Loh},\ and\ \citenamefont
  {Hirsch}}]{Scalapino1986}%
  \BibitemOpen
  \bibfield  {author} {\bibinfo {author} {\bibfnamefont {D.~J.}\ \bibnamefont
  {Scalapino}}, \bibinfo {author} {\bibfnamefont {E.}~\bibnamefont {Loh}}, \
  and\ \bibinfo {author} {\bibfnamefont {J.~E.}\ \bibnamefont {Hirsch}},\
  }\href {\doibase 10.1103/PhysRevB.34.8190} {\bibfield  {journal} {\bibinfo
  {journal} {Phys. Rev. B}\ }\textbf {\bibinfo {volume} {34}},\ \bibinfo
  {pages} {8190} (\bibinfo {year} {1986})}\BibitemShut {NoStop}%
\bibitem [{\citenamefont {Kopietz}\ \emph {et~al.}(2010)\citenamefont
  {Kopietz}, \citenamefont {Bartosch},\ and\ \citenamefont
  {Sch\"{u}tz}}]{Kopietz2010}%
  \BibitemOpen
  \bibfield  {author} {\bibinfo {author} {\bibfnamefont {P.}~\bibnamefont
  {Kopietz}}, \bibinfo {author} {\bibfnamefont {L.}~\bibnamefont {Bartosch}}, \
  and\ \bibinfo {author} {\bibfnamefont {F.}~\bibnamefont {Sch\"{u}tz}},\
  }\href {http://www.springer.com/de/book/9783642050930} {\emph {\bibinfo
  {title} {Introduction to the Functional Renormalization Group (Lecture Notes
  in Physics)}}}\ (\bibinfo  {publisher} {Springer, Berlin},\ \bibinfo {year}
  {2010})\BibitemShut {NoStop}%
\bibitem [{\citenamefont {Metzner}\ \emph {et~al.}(2012)\citenamefont
  {Metzner}, \citenamefont {Salmhofer}, \citenamefont {Honerkamp},
  \citenamefont {Meden},\ and\ \citenamefont {Sch\"onhammer}}]{Metzner2012}%
  \BibitemOpen
  \bibfield  {author} {\bibinfo {author} {\bibfnamefont {W.}~\bibnamefont
  {Metzner}}, \bibinfo {author} {\bibfnamefont {M.}~\bibnamefont {Salmhofer}},
  \bibinfo {author} {\bibfnamefont {C.}~\bibnamefont {Honerkamp}}, \bibinfo
  {author} {\bibfnamefont {V.}~\bibnamefont {Meden}}, \ and\ \bibinfo {author}
  {\bibfnamefont {K.}~\bibnamefont {Sch\"onhammer}},\ }\href {\doibase
  10.1103/RevModPhys.84.299} {\bibfield  {journal} {\bibinfo  {journal} {Rev.
  Mod. Phys.}\ }\textbf {\bibinfo {volume} {84}},\ \bibinfo {pages} {299}
  (\bibinfo {year} {2012})}\BibitemShut {NoStop}%
\bibitem [{\citenamefont {Platt}\ \emph {et~al.}(2013)\citenamefont {Platt},
  \citenamefont {Hanke},\ and\ \citenamefont {Thomale}}]{Platt2013}%
  \BibitemOpen
  \bibfield  {author} {\bibinfo {author} {\bibfnamefont {C.}~\bibnamefont
  {Platt}}, \bibinfo {author} {\bibfnamefont {W.}~\bibnamefont {Hanke}}, \ and\
  \bibinfo {author} {\bibfnamefont {R.}~\bibnamefont {Thomale}},\ }\href
  {\doibase 10.1080/00018732.2013.862020} {\bibfield  {journal} {\bibinfo
  {journal} {Advances in Physics}\ }\textbf {\bibinfo {volume} {62}},\ \bibinfo
  {pages} {453} (\bibinfo {year} {2013})},\ \Eprint
  {http://arxiv.org/abs/https://doi.org/10.1080/00018732.2013.862020}
  {https://doi.org/10.1080/00018732.2013.862020} \BibitemShut {NoStop}%
\bibitem [{\citenamefont {Husemann}\ and\ \citenamefont
  {Salmhofer}(2009)}]{Husemann2009}%
  \BibitemOpen
  \bibfield  {author} {\bibinfo {author} {\bibfnamefont {C.}~\bibnamefont
  {Husemann}}\ and\ \bibinfo {author} {\bibfnamefont {M.}~\bibnamefont
  {Salmhofer}},\ }\href {\doibase 10.1103/PhysRevB.79.195125} {\bibfield
  {journal} {\bibinfo  {journal} {Phys. Rev. B}\ }\textbf {\bibinfo {volume}
  {79}},\ \bibinfo {pages} {195125} (\bibinfo {year} {2009})}\BibitemShut
  {NoStop}%
\bibitem [{\citenamefont {Husemann}\ \emph {et~al.}(2012)\citenamefont
  {Husemann}, \citenamefont {Giering},\ and\ \citenamefont
  {Salmhofer}}]{Husemann2012}%
  \BibitemOpen
  \bibfield  {author} {\bibinfo {author} {\bibfnamefont {C.}~\bibnamefont
  {Husemann}}, \bibinfo {author} {\bibfnamefont {K.-U.}\ \bibnamefont
  {Giering}}, \ and\ \bibinfo {author} {\bibfnamefont {M.}~\bibnamefont
  {Salmhofer}},\ }\href {\doibase 10.1103/PhysRevB.85.075121} {\bibfield
  {journal} {\bibinfo  {journal} {Phys. Rev. B}\ }\textbf {\bibinfo {volume}
  {85}},\ \bibinfo {pages} {075121} (\bibinfo {year} {2012})}\BibitemShut
  {NoStop}%
\bibitem [{\citenamefont {Eberlein}(2015)}]{Eberlein2015}%
  \BibitemOpen
  \bibfield  {author} {\bibinfo {author} {\bibfnamefont {A.}~\bibnamefont
  {Eberlein}},\ }\href {\doibase 10.1103/PhysRevB.92.235146} {\bibfield
  {journal} {\bibinfo  {journal} {Phys. Rev. B}\ }\textbf {\bibinfo {volume}
  {92}},\ \bibinfo {pages} {235146} (\bibinfo {year} {2015})}\BibitemShut
  {NoStop}%
\bibitem [{\citenamefont {Wang}\ \emph {et~al.}(2012)\citenamefont {Wang},
  \citenamefont {Xiang}, \citenamefont {Wang}, \citenamefont {Wang},
  \citenamefont {Yang},\ and\ \citenamefont {Lee}}]{Wang2012}%
  \BibitemOpen
  \bibfield  {author} {\bibinfo {author} {\bibfnamefont {W.-S.}\ \bibnamefont
  {Wang}}, \bibinfo {author} {\bibfnamefont {Y.-Y.}\ \bibnamefont {Xiang}},
  \bibinfo {author} {\bibfnamefont {Q.-H.}\ \bibnamefont {Wang}}, \bibinfo
  {author} {\bibfnamefont {F.}~\bibnamefont {Wang}}, \bibinfo {author}
  {\bibfnamefont {F.}~\bibnamefont {Yang}}, \ and\ \bibinfo {author}
  {\bibfnamefont {D.-H.}\ \bibnamefont {Lee}},\ }\href {\doibase
  10.1103/PhysRevB.85.035414} {\bibfield  {journal} {\bibinfo  {journal} {Phys.
  Rev. B}\ }\textbf {\bibinfo {volume} {85}},\ \bibinfo {pages} {035414}
  (\bibinfo {year} {2012})}\BibitemShut {NoStop}%
\bibitem [{\citenamefont {Wang}\ \emph {et~al.}(2013)\citenamefont {Wang},
  \citenamefont {Li}, \citenamefont {Xiang},\ and\ \citenamefont
  {Wang}}]{Wang2013}%
  \BibitemOpen
  \bibfield  {author} {\bibinfo {author} {\bibfnamefont {W.-S.}\ \bibnamefont
  {Wang}}, \bibinfo {author} {\bibfnamefont {Z.-Z.}\ \bibnamefont {Li}},
  \bibinfo {author} {\bibfnamefont {Y.-Y.}\ \bibnamefont {Xiang}}, \ and\
  \bibinfo {author} {\bibfnamefont {Q.-H.}\ \bibnamefont {Wang}},\ }\href
  {\doibase 10.1103/PhysRevB.87.115135} {\bibfield  {journal} {\bibinfo
  {journal} {Phys. Rev. B}\ }\textbf {\bibinfo {volume} {87}},\ \bibinfo
  {pages} {115135} (\bibinfo {year} {2013})}\BibitemShut {NoStop}%
\bibitem [{\citenamefont {Lichtenstein}\ \emph {et~al.}(2017)\citenamefont
  {Lichtenstein}, \citenamefont {de~la Peña}, \citenamefont {Rohe},
  \citenamefont {Napoli}, \citenamefont {Honerkamp},\ and\ \citenamefont
  {Maier}}]{Lichtenstein2017}%
  \BibitemOpen
  \bibfield  {author} {\bibinfo {author} {\bibfnamefont {J.}~\bibnamefont
  {Lichtenstein}}, \bibinfo {author} {\bibfnamefont {D.~S.}\ \bibnamefont
  {de~la Peña}}, \bibinfo {author} {\bibfnamefont {D.}~\bibnamefont {Rohe}},
  \bibinfo {author} {\bibfnamefont {E.~D.}\ \bibnamefont {Napoli}}, \bibinfo
  {author} {\bibfnamefont {C.}~\bibnamefont {Honerkamp}}, \ and\ \bibinfo
  {author} {\bibfnamefont {S.}~\bibnamefont {Maier}},\ }\href {\doibase
  https://doi.org/10.1016/j.cpc.2016.12.013} {\bibfield  {journal} {\bibinfo
  {journal} {Computer Physics Communications}\ }\textbf {\bibinfo {volume}
  {213}},\ \bibinfo {pages} {100 } (\bibinfo {year} {2017})}\BibitemShut
  {NoStop}%
\bibitem [{\citenamefont {Schober}\ \emph {et~al.}(2018)\citenamefont
  {Schober}, \citenamefont {Ehrlich}, \citenamefont {Reckling},\ and\
  \citenamefont {Honerkamp}}]{Schober2018}%
  \BibitemOpen
  \bibfield  {author} {\bibinfo {author} {\bibfnamefont {G.~A.~H.}\
  \bibnamefont {Schober}}, \bibinfo {author} {\bibfnamefont {J.}~\bibnamefont
  {Ehrlich}}, \bibinfo {author} {\bibfnamefont {T.}~\bibnamefont {Reckling}}, \
  and\ \bibinfo {author} {\bibfnamefont {C.}~\bibnamefont {Honerkamp}},\ }\href
  {\doibase 10.3389/fphy.2018.00032} {\bibfield  {journal} {\bibinfo  {journal}
  {Frontiers in Physics}\ }\textbf {\bibinfo {volume} {6}},\ \bibinfo {pages}
  {32} (\bibinfo {year} {2018})}\BibitemShut {NoStop}%
\bibitem [{\citenamefont {de~la Pe\~na}\ \emph {et~al.}(2017)\citenamefont
  {de~la Pe\~na}, \citenamefont {Lichtenstein},\ and\ \citenamefont
  {Honerkamp}}]{Sanchez2017}%
  \BibitemOpen
  \bibfield  {author} {\bibinfo {author} {\bibfnamefont {D.~S.}\ \bibnamefont
  {de~la Pe\~na}}, \bibinfo {author} {\bibfnamefont {J.}~\bibnamefont
  {Lichtenstein}}, \ and\ \bibinfo {author} {\bibfnamefont {C.}~\bibnamefont
  {Honerkamp}},\ }\href {\doibase 10.1103/PhysRevB.95.085143} {\bibfield
  {journal} {\bibinfo  {journal} {Phys. Rev. B}\ }\textbf {\bibinfo {volume}
  {95}},\ \bibinfo {pages} {085143} (\bibinfo {year} {2017})}\BibitemShut
  {NoStop}%
\bibitem [{\citenamefont {Vilardi}\ \emph {et~al.}(2017)\citenamefont
  {Vilardi}, \citenamefont {Taranto},\ and\ \citenamefont
  {Metzner}}]{Vilardi2017}%
  \BibitemOpen
  \bibfield  {author} {\bibinfo {author} {\bibfnamefont {D.}~\bibnamefont
  {Vilardi}}, \bibinfo {author} {\bibfnamefont {C.}~\bibnamefont {Taranto}}, \
  and\ \bibinfo {author} {\bibfnamefont {W.}~\bibnamefont {Metzner}},\ }\href
  {\doibase 10.1103/PhysRevB.96.235110} {\bibfield  {journal} {\bibinfo
  {journal} {Phys. Rev. B}\ }\textbf {\bibinfo {volume} {96}},\ \bibinfo
  {pages} {235110} (\bibinfo {year} {2017})}\BibitemShut {NoStop}%
\bibitem [{\citenamefont {Tagliavini}\ \emph {et~al.}(2019)\citenamefont
  {Tagliavini}, \citenamefont {Hille}, \citenamefont {Kugler}, \citenamefont
  {Andergassen}, \citenamefont {Toschi},\ and\ \citenamefont
  {Honerkamp}}]{Tagliavini2019}%
  \BibitemOpen
  \bibfield  {author} {\bibinfo {author} {\bibfnamefont {A.}~\bibnamefont
  {Tagliavini}}, \bibinfo {author} {\bibfnamefont {C.}~\bibnamefont {Hille}},
  \bibinfo {author} {\bibfnamefont {F.~B.}\ \bibnamefont {Kugler}}, \bibinfo
  {author} {\bibfnamefont {S.}~\bibnamefont {Andergassen}}, \bibinfo {author}
  {\bibfnamefont {A.}~\bibnamefont {Toschi}}, \ and\ \bibinfo {author}
  {\bibfnamefont {C.}~\bibnamefont {Honerkamp}},\ }\href {\doibase
  10.21468/SciPostPhys.6.1.009} {\bibfield  {journal} {\bibinfo  {journal}
  {SciPost Phys.}\ }\textbf {\bibinfo {volume} {6}},\ \bibinfo {pages} {9}
  (\bibinfo {year} {2019})}\BibitemShut {NoStop}%
\bibitem [{\citenamefont {Salmhofer}\ and\ \citenamefont
  {Honerkamp}(2001)}]{Salmhofer2001}%
  \BibitemOpen
  \bibfield  {author} {\bibinfo {author} {\bibfnamefont {M.}~\bibnamefont
  {Salmhofer}}\ and\ \bibinfo {author} {\bibfnamefont {C.}~\bibnamefont
  {Honerkamp}},\ }\href {\doibase 10.1143/PTP.105.1} {\bibfield  {journal}
  {\bibinfo  {journal} {Progress of Theoretical Physics}\ }\textbf {\bibinfo
  {volume} {105}},\ \bibinfo {pages} {1} (\bibinfo {year} {2001})},\ \Eprint
  {http://arxiv.org/abs/http://ptp.oxfordjournals.org/content/105/1/1.full.pdf+html}
  {http://ptp.oxfordjournals.org/content/105/1/1.full.pdf+html} \BibitemShut
  {NoStop}%
\bibitem [{\citenamefont {Kugler}\ and\ \citenamefont {von
  Delft}(2018)}]{Kugler2018}%
  \BibitemOpen
  \bibfield  {author} {\bibinfo {author} {\bibfnamefont {F.~B.}\ \bibnamefont
  {Kugler}}\ and\ \bibinfo {author} {\bibfnamefont {J.}~\bibnamefont {von
  Delft}},\ }\href {\doibase 10.1103/PhysRevLett.120.057403} {\bibfield
  {journal} {\bibinfo  {journal} {Phys. Rev. Lett.}\ }\textbf {\bibinfo
  {volume} {120}},\ \bibinfo {pages} {057403} (\bibinfo {year}
  {2018})}\BibitemShut {NoStop}%
\bibitem [{\citenamefont {Scalettar}\ \emph {et~al.}(1989)\citenamefont
  {Scalettar}, \citenamefont {Scalapino}, \citenamefont {Sugar},\ and\
  \citenamefont {Toussaint}}]{Scalettar1989}%
  \BibitemOpen
  \bibfield  {author} {\bibinfo {author} {\bibfnamefont {R.~T.}\ \bibnamefont
  {Scalettar}}, \bibinfo {author} {\bibfnamefont {D.~J.}\ \bibnamefont
  {Scalapino}}, \bibinfo {author} {\bibfnamefont {R.~L.}\ \bibnamefont
  {Sugar}}, \ and\ \bibinfo {author} {\bibfnamefont {D.}~\bibnamefont
  {Toussaint}},\ }\href {\doibase 10.1103/PhysRevB.39.4711} {\bibfield
  {journal} {\bibinfo  {journal} {Phys. Rev. B}\ }\textbf {\bibinfo {volume}
  {39}},\ \bibinfo {pages} {4711} (\bibinfo {year} {1989})}\BibitemShut
  {NoStop}%
\bibitem [{\citenamefont {Kanamori}(1963)}]{Kanamori1963}%
  \BibitemOpen
  \bibfield  {author} {\bibinfo {author} {\bibfnamefont {J.}~\bibnamefont
  {Kanamori}},\ }\href {\doibase 10.1143/PTP.30.275} {\bibfield  {journal}
  {\bibinfo  {journal} {Progress of Theoretical Physics}\ }\textbf {\bibinfo
  {volume} {30}},\ \bibinfo {pages} {275} (\bibinfo {year} {1963})},\ \Eprint
  {http://arxiv.org/abs/http://oup.prod.sis.lan/ptp/article-pdf/30/3/275/5278869/30-3-275.pdf}
  {http://oup.prod.sis.lan/ptp/article-pdf/30/3/275/5278869/30-3-275.pdf}
  \BibitemShut {NoStop}%
\bibitem [{\citenamefont {{J\"{u}lich Supercomputing Centre}}(2018)}]{jureca}%
  \BibitemOpen
  \bibfield  {author} {\bibinfo {author} {\bibnamefont {{J\"{u}lich
  Supercomputing Centre}}},\ }\href {\doibase 10.17815/jlsrf-4-121-1}
  {\bibfield  {journal} {\bibinfo  {journal} {Journal of large-scale research
  facilities}\ }\textbf {\bibinfo {volume} {4}} (\bibinfo {year} {2018}),\
  10.17815/jlsrf-4-121-1}\BibitemShut {NoStop}%
\end{thebibliography}%

\end{document}